\documentclass[lettersize,journal]{IEEEtran}
\usepackage{amsmath,amsfonts,amsthm}
\usepackage{algorithmic}
\usepackage{algorithm}
\usepackage{array}
\usepackage{enumitem}
\usepackage{mathtools}
\usepackage{dsfont}
\usepackage{textcomp}
\usepackage{tikz}
\usepackage{stfloats}
\usepackage{url}
\usepackage{verbatim}
\usepackage{xcolor, color, colortbl}
\usepackage{graphicx}
\usepackage{cite}
\usepackage[T1]{fontenc}
\usepackage[utf8]{inputenc}
\usepackage{xpatch}
\usepackage{fancyhdr}

\makeatletter
\let\MYcaption\@makecaption
\makeatother
\usepackage[font=footnotesize]{subcaption}
\makeatletter
\let\@makecaption\MYcaption
\makeatother

\graphicspath{{fig/}}

\makeatletter
\xpatchcmd{\@thm}{\thm@headpunct{.}}{\thm@headpunct{}}{}{}
\makeatother

\makeatother
\newtheorem{theorem}{Theorem}
\newtheorem{lemma}{Lemma}

\newtheorem{assumption}{Assumption}
\newtheorem{remark}{Remark}

\newtheorem{condition}{Condition}

\newcommand*{\C}{\mathbb{C}}
\newcommand*{\R}{\mathbb{R}}

\DeclareMathOperator{\diff}{d}
\DeclareMathOperator{\diag}{diag}
\newcommand{\ddt}{\tfrac{\diff}{\diff \!t}}
\definecolor{light-gray}{gray}{0.96}

\DeclareMathAlphabet{\mymathbb}{U}{bbold}{m}{n}

\definecolor{dwSG1}{rgb}{0.00000,0.44700,0.74100}%
\definecolor{dwSG2}{rgb}{0.85000,0.32500,0.09800}%
\definecolor{dwSG3}{rgb}{0.92900,0.69400,0.12500}%

\begin{document}


\title{Certifying Frequency Stability for Systems with Line Dynamics and Heterogeneous Bus Dynamics}

\author{Dahlia~Saba, Dominic~Gro\ss \vspace{-1em}%
\thanks{D. Saba and D. Groß are with the Department of
Electrical and Computer Engineering at the University of Wisconsin-Madison,
USA; e-mail:{dsaba, dominic.gross}@wisc.edu}
\thanks{Manuscript received January 26, 2026; revised July 17, 2026.}}



\maketitle

\begin{abstract}
    This work presents a framework for certifying small-signal frequency stability of a power system with line dynamics and heterogeneous bus dynamics. This framework can certify the  stability of systems which include synchronous generators, synchronous condensers, and converter-interfaced resources with a wide range of controls. Moreover, it can do so without detailed or precise knowledge of the network topology. With this framework, we also provide a detailed analysis of how proportional-derivative (PD) droop can improve the stability margin of the frequency response. The stability certificates presented in this work, which extend prior results by incorporating line dynamics, provide insight into how the control parameters for different units in the system impact the overall frequency stability.
    While damper windings have long been understood to improve the frequency synchronization between machines, the dynamics of the damper windings are complex, making them difficult to analyze. To address this gap, this paper derives a novel reduced-order model of the damper windings in the form of a derivative droop term. Moreover, we show that derivative droop terms used in grid-forming (GFM) control can be understood as a form of damper winding emulation. Our analytical stability conditions highlight the importance of damper windings (or their emulation) in facilitating frequency synchronization and suppressing unstable interactions between GFM converters. These results are validated with electromagnetic-transient (EMT) simulation.
\end{abstract}

%

\vspace{-1em}

\section{Introduction}
\IEEEPARstart{C}{onverter}-interfaced resources such as wind, solar photovoltaics, and battery energy storage have been rapidly deployed to bulk power systems. Existing deployments overwhelmingly use voltage source converters (VSCs) with grid-following (GFL) control, which relies on the presence of conventional synchronous generators to ensure grid stability. Grid-forming (GFM) control has been proposed as an alternative that allows converter-based resources to operate grids without synchronous generators~\cite{matevosyan_grid-forming_2019, lasseter_grid-forming_2020}. Historically, stability analysis of bulk power systems has relied on the physical properties of synchronous machines, namely their inherently high rotational inertia and well-understood controls. Converter-interfaced resources lack these physical properties: VSCs have limited internal energy storage, giving them faster dynamics and lower interia, and many more degrees of freedom in their controls. As a result, new analysis methods are needed to ensure the stability of bulk power systems that contain both synchronous machines and converter-based resources. 

Methods for certifying the stability of power systems with converter-interfaced resources in the literature generally fall into three categories: large-signal analytical conditions, small-signal analytical conditions, and small-signal numerical conditions. 
Large-signal stability criteria certify stability of the nonlinear power system dynamics, but typically only apply to networks of homogeneous converters (e.g., \cite{gros_effect_2019}). On the other hand, numerical small-signal  analysis (e.g., impedance models~\cite{cespedes_impedance_2014,harnefors_passivity-based_2016,yu_analysis_2016,liao_sub-synchronous_2019} and eigenvalue sensitivity analysis~\cite{pogaku_modeling_2007}), which account for  heterogeneous controls and circuit dynamics, require modeling the entire power system under study. While impedance methods have been widely used for tuning converter inner loops and assessing harmonic stability, their model requirements and computational complexity often makes them impractical for analyzing large networks and results often do not generalize to systems beyond the specific system under study. 


Existing bus-level analytical small-signal stability conditions for networks of converter-interfaced resources with line dynamics are highly scalable~\cite{vorobev_framework_2017,gross_compensating_2022,bui_input-output_2024,haberle_decentralized_2025}. Bus-level stability criteria have been derived for networks with droop-controlled converters in microgrids with heterogeneous resistance-to-reactance ($R/X$) ratios~\cite{vorobev_framework_2017}, droop-controlled converters in transmission networks with homogeneous line dynamics~\cite{haberle_decentralized_2025}, and for proportional-derivative droop and frequency and angle droop~\cite{gross_compensating_2022}. These results all assume the network is solely composed of converters limited to one type of control, and thus are unable to account for instabilities that may arise in heterogeneous networks (e.g., interactions between a converter and a synchronous generator).

Fewer analytical works exist in the literature that certify stability for a network with heterogeneous bus dynamics that are not restricted to any particular control. A set of bus-level (i.e., decentralized) stability criteria is presented in~\cite{pates_robust_2019}, and~\cite{bui_input-output_2024} extends its results to account for the effects of line dynamics with a homogeneous $R/X$ ratio.
However, both of these results do not cover marginally stable bus dynamics, which rules out common devices such as synchronous condensers, static synchronous compensators (STATCOMs), and many renewables operating at their maximum power point. In contrast, the conditions in~\cite{saba_frequency_2025} allow for marginally stable bus dynamics and a reduced-order model damper winding model is introduced to demonstrate that synchronous condensers meet these conditions. However, the results of~\cite{saba_frequency_2025} do not cover line dynamics.

In this work, we develop a framework for certifying small-signal frequency stability of a power system with line dynamics and heterogeneous bus dynamics that relies on bus-level conditions and a few key system-level conditions. In contrast to conventional eigenvalue sensitivity and impedance methods, the stability conditions proposed in this work do not require a full model of either the converters or the network, which are difficult to obtain at a system level in practice. Moreover, because our framework is analytical and decentralized, it can provide generalizable insight about how to design and tune the bus controls, unlike the conventional numerical methods.

In particular, our stability framework provides insights into the effect of machine damper windings as a form of active damping that helps compensate the effects of the line dynamics. One significant difference between synchronous machines and converter-based resources is the presence of damper windings in machines, which help damp rotor oscillations~\cite{ebrahimi_improved_2019, zhao_transient_2023}. Existing works have proposed designing converter controls to emulate the effect of damper windings. One approach is to approximate the induced damping power with the difference between the estimated grid frequency and the converter frequency. Some works accomplish this by measuring the grid frequency with a phase-locked loop (PLL)~\cite{gao_control_2008,shintai_oscillation_2014}; however, the PLL can cause instability, particularly in weak grids~\cite{dong_analysis_2015}. Other work adds a frequency damping term that effectively approximates the grid frequency with the nominal frequency~\cite{zhong_synchronverters_2011}, but this model is inaccurate, leading to undesirable behavior if the grid frequency is not at its nominal value. Alternate approaches emulate the effects of damper windings without estimating the grid frequency by adjusting the converter duty cycle~\cite{cvetkovic_modeling_2015}, adjusting the output voltage amplitude~\cite{ebrahimi_improved_2019}, adding virtual flux variables~\cite{yin_attenuation_2022}, or using a transient virtual impedance~\cite{gajare_grid-forming_2025}. However, the system-level impact of these controls is difficult to mathematically analyze. Therefore, while~\cite{cvetkovic_modeling_2015,ebrahimi_improved_2019,yin_attenuation_2022,gajare_grid-forming_2025} demonstrate stability benefits in small numerical studies, it is unclear how to tune these controls or if these effects generalize to larger systems.

This paper provides a more rigorous justification of the reduced-order damper winding model in~\cite{saba_frequency_2025}, showing that the effect of damper windings can be described by a proportional-derivative (PD) term, and extends it to a network with line dynamics. 
We further show that emulating the damper winding model by adding a derivative droop term to GFM controls similarly improves the dynamic performance of the converter, helping counteract the destabilizing effects of line dynamics. This provides physical intuition for the results of~\cite{gross_compensating_2022}, which found that PD converter controls can compensate line dynamics in a system with homogeneous bus dynamics, and extends those results to systems with heterogeneous and marginally stable bus dynamics. While PD droop control (also referred to as lead-lag control) has been proposed before~\cite{engler_device_2006,guerrero_decentralized_2007,mohamed_adaptive_2008,guerrero_control_2009}, the analytical stability framework presented in this work provides concrete guidelines for how to tune the controls. Furthermore, this work provides an interpretation of derivative droop control as a form of damper winding emulation that, in contrast to other controls mimicking damper windings, is simple to implement and is shown to help stabilize system-level frequency dynamics even for large systems with heterogeneous bus dynamics. 

The main contributions of this work are as follows: \begin{enumerate}
    \item We introduce scalable analytical system-level and bus-level conditions for certifying frequency stability of a power system with line dynamics and heterogeneous bus dynamics that allows for marginally stable bus dynamics (e.g., synchronous condensers and STATCOMs).
    \item We derive a reduced-order damper winding model based on approximating the full order synchronous machine dynamics near steady state. This model improves upon the accuracy of the conventional swing equation model by accounting for the effects of damper windings that are conventionally neglected. The accuracy improvement is validated numerically against an electromagnetic-transient (EMT) simulation.
    \item We revisit PD droop control and show that the derivative term can be understood as a form of damper winding emulation. This result directly extends to dual-port GFM control.
    \item Using the scalable analytical stability conditions, we analyze the impact of derivative droop in GFM converters and synchronous machines. We illustrate how derivative droop can compensate the destabilizing effects of line dynamics. This analysis demonstrates the importance of accounting for the effects of machine damper windings when analyzing dynamic stability and provides practical insight into how to tune the control parameters of PD droop controlled converters.
\end{enumerate}

The rest of the paper follows the following structure: Section~\ref{sec:dw_derivation} derives the reduced-order damper winding model from the synchronous machine dynamics. With this reduced-order model, Sec.~\ref{sec:model} presents a small-signal model of the power grid which accounts for line dynamics. Section~\ref{sec:stability} derives a set of decentralized stability conditions to certify small-signal stability of the power system model. Section~\ref{sec:analyze_dw_effects} uses the stability criteria and reduced-order damper winding model to analyze how accounting for the effects of damper windings helps demonstrate the stability of synchronous machines in the presence of line dynamics. Section~\ref{sec:dw_emulation} then uses the stability framework from Sec.~\ref{sec:stability} to demonstrate how augmenting conventional GFM droop control with a PD term to implement damper winding emulation can provide stability benefits. Finally, Sec.~\ref{sec:simulation} validates the analytical stability results through an EMT simulation of the IEEE 9-bus system~\cite[p. 38]{paul_m_anderson_power_1977} with two synchronous generators and a GFM converter plant.

\section{Reduced-Order Damping Torque Model}\label{sec:dw_derivation}

\begin{figure}
    \centering
    \includegraphics[width=0.8\columnwidth]{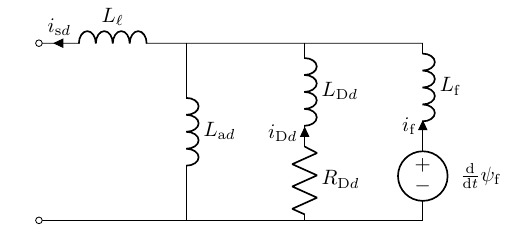}
    \caption{Model of the $d$-axis synchronous machine circuit, where $L_{\ell}$ denotes the stator leakage inductance, $L_{\text{a}d}$ denotes the $d$-axis magnetizing inductance, $L_{\text{D}d}$ and $R_{\text{D}d}$ denote the $d$-axis damper winding inductance and resistance, and $L_{\text{f}}$ and $\psi_{\text{f}}$ denote the field winding inductance and flux linkage. The $q$-axis circuit is modeled by replacing the field winding with an open circuit.}\label{fig:sg_dynamic_circuit}
    \vspace{-1em}
\end{figure}
%
%
Before presenting our power system model (see Sec.~\ref{sec:model}), we first introduce a reduce-order model of the machine damper windings that facilitates our stability analysis of multi-machine/multi-converter systems. In this section, we derive a reduced-order model of the impact of synchronous machine damper windings on the electrical torque near steady state. Let the electrical torque $\tau_\text{e} \in \R$ be expressed as 
\begin{align}
    \tau_{\text{e}} = \tau_{\text{net}} + \tau_\text{D},\label{eq:tau_e}
\end{align}
where $\tau_{\text{net}} \in \R$ represents the torque corresponding to the active power injection into the grid and $\tau_{\text{D}} \in \R$ represents the torque induced by the damper windings. We show that the damping torque can be approximated by 
    \begin{align}
        \tau_{\text{D}} =  \xi_{\text{SM}} \ddt \tau_{\text{net}}.\label{eq:damping_torque}
    \end{align}
This derivation is based on the synchronous machine dynamic model presented in~\cite[Ch. 3, Ch. 5.3.2]{kundur_power_1994} and improves on the result from~\cite[Sec. 3]{saba_frequency_2025}. In contrast to~\cite[Sec. 3]{saba_frequency_2025}, which assumes a multi-machine network in quasi-steady state without converters, this derivation applies to a machine connected to an arbitrary network which may have line dynamics. 

Consider a synchronous machine with rotor angle $\theta_\text{r} \in [0,\, 2\pi)$ and a $d$-axis and a $q$-axis damper winding
. It is assumed that the Canay inductance, stator resistance, and field winding resistance can be neglected. The equivalent circuit is shown in Fig.~\ref{fig:sg_dynamic_circuit}. Assuming all three-phase voltages and currents are balanced, we define a rotating $dq$-frame aligned with $\theta_\text{r}$. Moreover, we define the $90^{\circ}$ rotation matrix as \[\mathbf{j} = \begin{bmatrix} 0 & -1 \\ 1 & 0 \end{bmatrix}.\] 

Let $\psi_\text{a} = [\psi_{\text{a}d},\,\psi_{\text{a}q}]^{\top} \in \R^2$ denote the air-gap flux linkage and $i_\text{s} = [i_{\text{s}d},\, i_{\text{s}q}]^{\top} \in \R^2$ denote the stator current in $dq$-frame. Moreover, let $L_{\ell} \in \R_{\geq 0}$ denote the stator leakage inductance. The electrical torque $\tau_\text{e} \in \R$ can be expressed as 
\begin{align}
    \tau_\text{e} = i_\text{s}^{\top}\mathbf{j}(\psi_\text{a} - L_{\ell}i_\text{s}) = i_\text{s}^{\top}\mathbf{j}\psi_\text{a}.\label{eq:torque}
\end{align}
Let $\psi_\text{D} = [\psi_{\text{D}d},\, \psi_{\text{D}q}]^{\top} \in \R^2$ denote the vector of flux linkages through the $d$- and $q$-axis damper windings and $\psi_\text{f} \in \R$ denote the field winding flux linkage. Likewise, let $L_{\text{a}d}, L_{\text{a}q} \in \R_{>0}$ denote the $d$- and $q$-axis magnetizing inductances, $L_{\text{D}d}, L_{\text{D}q} \in \R_{>0}$ denote the $d$- and $q$-axis damper winding inductances, and $L_\text{f} \in \R_{>0}$ denote the field winding inductance. The air gap flux linkage is given by $\psi_{\text{a}} = -L_{\text{a}}\left(-i_{\text{s}} + i_{\text{D}} + [i_{\text{f}},\, 0]^{\top}\right)$, where $i_\text{f} \in \R$ is the field winding current. Moreover, the field winding flux linkage can be expressed as $\psi_{\text{f}} = \psi_{\text{a},d} + L_{\text{f}}i_\text{f}$, and the damper winding flux linkage can be expressed as $\psi_{\text{D}} = \psi_{\text{a}} + L_{\text{D}}i_\text{D}$~\cite[p. 188]{kundur_power_1994}. By defining the subtransient inductances $L_{\text{a}d}^{\prime\prime} \coloneqq (L_{\text{a}d}^{-1} + L_\text{f}^{-1}+L_{\text{D}d}^{-1})^{-1}$ and $L_{\text{a}q}^{\prime\prime} \coloneqq (L_{\text{a}q}^{-1} +L_{\text{D}q}^{-1})^{-1}$, the air-gap flux linkage can be expressed as
\begin{align}
    \psi_{\text{a}} = L_\text{a}^{\prime\prime}\left(-i_\text{s} + L_\text{D}^{-1}\psi_\text{D} + \frac{1}{L_\text{f}}\begin{bmatrix}\psi_\text{f} \\ 0\end{bmatrix}\right), \label{eq:stator_flux}
\end{align}
where $L_\text{a}^{\prime\prime} \coloneqq \diag(L_{\text{a}d}^{\prime\prime}, L_{\text{a}q}^{\prime\prime}) \in \R^{2\times 2}$ and $L_\text{D} \coloneqq \diag(L_{\text{D}d},\, L_{\text{D}q}) \in \R^{2\times 2}$. Substituting~\eqref{eq:stator_flux} into~\eqref{eq:torque}, we can express the electrical torque in the form of~\eqref{eq:tau_e} as 
\begin{align}\label{eq:trq_split}
    \tau_\text{e} = \underbrace{\frac{L_{\text{a}d}^{\prime\prime}}{L_\text{f}} i_{\text{s}q} \psi_\text{f}+ (L_{\text{a}q}^{\prime\prime} - L_{\text{a}d}^{\prime\prime})i_{\text{s}d}i_{\text{s}q}}_{\tau_{\text{net}}} + \underbrace{i_\text{s}^{\top}\mathbf{j}L_\text{a}^{\prime\prime}L_\text{D}^{-1}\psi_\text{D}}_{\tau_\text{D}}.
\end{align}

To solve for $\psi_\text{D}$, we first solve for the damper winding currents, represented in $dq$ coordinates by $i_\text{D} = [i_{
    \text{D}d}, \, i_{\text{D}q}]^{\top} \in \R^{2}$. The damper currents are given by 
\begin{align}
    R_\text{D} i_\text{D} = -\ddt \psi_\text{D},\label{eq:dw_dyn}
\end{align}
where $R_\text{D} = \diag(r_{\text{D}d},\, r_{\text{D}q}) \in \R^{2\times 2}$ is the matrix of damper winding resistances. 
Assuming the field winding flux is constant, substituting~\eqref{eq:stator_flux} into the expression for $\psi_\text{D}$ and taking the derivative of both sides results in
\begin{align*}
    \ddt \psi_\text{D} = L_\text{D} \ddt i_\text{D} + L_\text{a}^{\prime\prime} \left(-\ddt i_\text{s} + L_\text{D}^{-1}\ddt \psi_\text{D}\right), 
\end{align*}
which can be rewritten to obtain
\begin{align}
    \ddt \psi_\text{D} = (L_\text{D} - L_\text{a}^{\prime\prime})^{-1} L_\text{D} \left(L_\text{D} \ddt i_\text{D} - L_\text{a}^{\prime\prime}\ddt i_\text{s}\right).\label{eq:ddt_dw_flux}
\end{align}
The damper windings act like the bars of an induction machine. We can therefore express the dynamics of the damping current in terms of the difference between the electrical frequency of the rotor and stator such that $\ddt i_\text{D} = (\omega_\text{s} - \omega_\text{r})\mathbf{j}i_\text{D}$, where $\omega_\text{s} \in \R$ is the stator frequency and $\omega_\text{r} \in \R$ is the rotor frequency. Moreover, if we assume the stator current has constant magnitude, we can express its dynamics in the rotating frame as $\ddt i_\text{s} = (\omega_\text{s} - \omega_\text{r})\mathbf{j}i_\text{s}$. Using these expressions, we can substitute~\eqref{eq:ddt_dw_flux} into~\eqref{eq:dw_dyn} to obtain 
\begin{align*}
    R_\text{D} i_\text{D} = - (\omega_\text{s} - \omega_\text{r}) (L_\text{D} - L_\text{a}^{\prime\prime})^{-1} L_\text{D} \left(L_\text{D} \mathbf{j}i_\text{D} - L_\text{a}^{\prime\prime}\mathbf{j}i_\text{s}\right).
\end{align*}
Solving for $i_\text{D}$ and linearizing around $\omega_\text{s} = \omega_\text{r}$ results in
\begin{align}
    i_\text{D} = (\omega_\text{s} - \omega_\text{r})R_\text{D}^{-1} (L_\text{D} - L_\text{a}^{\prime\prime})^{-1}L_\text{D} L_\text{a}^{\prime\prime}\mathbf{j}i_\text{s}.
\end{align}
Neglecting any coupling between the field winding and damper winding fluxes, the damper winding flux linkage can now be expressed as 
\begin{align*}
    \psi_\text{D} &= - L_\text{a} i_\text{s} + L_\text{D} i_\text{D} \\
           &= \left(-L_\text{a} + (\omega_\text{s} - \omega_\text{r}) R_\text{D}^{-1} (L_\text{D} - L_\text{a}^{\prime\prime})^{-1}L_\text{D}^2 L_\text{a}^{\prime\prime}\mathbf{j}\right)i_\text{s}.
\end{align*}
Substituting this into the damping torque in \eqref{eq:trq_split} results in 
\begin{multline*}
    \tau_\text{D} = -i_\text{s}^{\top}\mathbf{j}L_\text{a}^{\prime\prime}L_\text{D}^{-1}L_\text{a} i_\text{s} + \\ i_\text{s}^{\top}\mathbf{j}(\omega_\text{s} - \omega_\text{r}) R_\text{D}^{-1} (L_\text{D} - L_\text{a}^{\prime\prime})^{-1}L_\text{D} {L_\text{a}^{\prime\prime}}^2\mathbf{j}i_\text{s}.
\end{multline*}
Because the stator is predominantly inductive and voltages magnitudes are assumed to be at their nominal value, the current will be close to $90^{\circ}$ out of phase with the rotor voltage in steady state conditions. As a result, we evaluate the expression at the point $i_\text{s} = [0, \, i_{\text{s}q}]^{\top}$, and $\tau_\text{D}$ reduces to 
\begin{align}
    \tau_\text{D} = -(\omega_\text{s} - \omega_\text{r}) \frac{L_{\text{D}d} {L_{\text{a}d}^{\prime\prime}}^2}{R_{\text{D}d}(L_{\text{D}d} - L_{\text{a}d}^{\prime\prime})}i_{\text{s}q}^2.
\end{align}

Now consider the derivative of $\tau_{\text{net}}$. Using $\ddt i_\text{s} = (\omega_\text{s} - \omega_\text{r}) \mathbf{j} i_\text{s}$, and assuming $\ddt \psi_\text{f} = 0$,  it holds that
\begin{align*}
    \ddt \tau_{\text{net}} = (\omega_\text{s} - \omega_\text{r})\left[L_{\text{a}d}^{\prime\prime}i_{\text{s}d} \psi_\text{f} + (L_{\text{a}q}^{\prime\prime} - L_{\text{a}d}^{\prime\prime})\left(i_{\text{s}d}^2 - i_{\text{s}q}^2\right)\right].
\end{align*}
We again evaluate this expression at the operating point $i_{\text{s}d} = 0$. This gives the desired result~\eqref{eq:damping_torque}, where $\xi_{\text{SM}}$ is given by
\begin{align}
    \xi_{\text{SM}} \coloneqq \frac{L_{\text{D}d} {L_{\text{a}d}^{\prime\prime}}^2}{R_{\text{D}d}(L_{\text{D}d} - L_{\text{a}d}^{\prime\prime})(L_{\text{a}q}^{\prime\prime}-L_{\text{a}d}^{\prime\prime})}.\label{eq:xi}
\end{align}
This model is validated via EMT simulation in Sec.~\ref{sec:sm_model}.

\section{Power System Model}\label{sec:model}
This section presents the model used to analyze the bulk power system. First, we discuss the advantage of using a polar coordinate frame for our system-level frequency stability analysis. Next, the network power flow model is introduced, including second-order line dynamics. Dynamic models are then presented for buses with converters and with synchronous machines, which are modeled as transfer functions from power to frequency. For synchronous machines, the transfer function incorporates the reduced-order damper winding model from Sec.~\ref{sec:dw_derivation}. Finally, the full frequency domain model of the interconnected system is introduced. 

\subsection{Coordinate Frame}

In this paper, we model the system dynamics in a power to frequency polar coordinate frame. This polar coordinate frame allows us to best capture the small-signal system response to slower dynamics, up to and including line frequency, and thus complements the $dq$ coordinate frame used in impedance stability analysis. In the $dq$ frame, the faster inner loop and filter dynamics are linear, which makes the $dq$ frame most useful for capturing the system response to faster disturbances (i.e., oscillations with frequency above line frequency). In constrast, in the polar coordinate frame the outer controls of converters (e.g., droop control) and the swing equation of synchronous machines are linear, making the polar coordinate frame more useful for capturing the system response to slower disturbances (i.e., oscillations up to line frequency). Note that because transformers and line impedances act as lowpass filters, the faster oscillations tend to remain local while these slower oscillations can propagate through the network, causing oscillations between different buses. Thus, to analyze the system-wide frequency stability, we focus on these slower oscillations and use the polar coordinate frame to model the system dynamics.

\subsection{Dynamic Small-signal Power Flow Model}
Let the transmission network be modeled by a graph where $\mathcal{N}_{\text{tot}}$ is the set of nodes corresponding to the buses and $\mathcal{E}_{\text{tot}} \subseteq \mathcal{N}_{\text{tot}} \times \mathcal{N}_{\text{tot}}$ is the set of edges corresponding to the transmission lines. Transmission lines are modeled as a series inductance and resistance. The dynamics of the current flowing across a line $(n,k) \in \mathcal{E}_{\text{tot}}$ are therefore described in $dq$-frame by
\begin{align*}
    \ell_{nk} \ddt \begin{bmatrix}
        i_{nk,d} \\ i_{nk,q}
    \end{bmatrix} = 
    \begin{bmatrix}
        -r_{nk} & \omega_0\ell_{nk} \\
        -\omega_0\ell_{nk} & -r_{nk}
    \end{bmatrix}
    \begin{bmatrix}
        i_{nk,d} \\ i_{nk,q}
    \end{bmatrix} + \begin{bmatrix}
        v_{n,d} - v_{k,d} \\
        v_{n,q} - v_{k,q}
    \end{bmatrix}
\end{align*}
where $i_{nk,d}$ and $i_{nk,q}$ are the $dq$ components of the line current, $v_{n,d}$ and $v_{n,q}$ are the $dq$ components of the voltage at bus $n$, $v_{k,d}$ and $v_{k,q}$ are the $dq$ components of the voltage at bus $k$, $r_{nk}$ and $\ell_{nk}$ are respectively the line's resistance and inductance, and $\omega_0$ is the nominal system frequency.

Let $V_n$ and $V_k$ be the magnitudes and $\theta_n$ and $\theta_k$ be the phase angles of the voltage waveforms at buses $n$ and $k$ respectively. By linearizing around the operating point $V_n = V_k = V^{\star}$ and $\theta_n = \theta_k$, we can express the dynamics of the power $p_{nk}$ flowing along the line $(n,k)$ in frequency domain as 
\begin{align}
    p_{nk}(s) = \frac{V^{{\star}2}}{\omega_0\ell_{nk}} \frac{\omega_0^2}{s^2 + 2\frac{r_{nk}}{\ell_{nk}}s + (\frac{r_{nk}}{\ell_{nk}})^2 + \omega_0^2}(\theta_{n}(s) - \theta_k(s)).\label{eq:p_nk_arbitrary_rx}
\end{align}

For simplicity of analysis, we assume all lines in the network have the same resistance-reactance ratio. While this assumption is not generally satisfied, it is a widely-used approximation that allows us to more easily model the effect of line dynamics. Moreover, because the differences in $R/X$ ratios on transmission systems are generally very small, they are generally negligible and the results of this analysis still generalize to more realistic networks with heterogeneous line dynamics (see Sec.~\ref{sec:simulation}). 
\begin{assumption}[\textbf{Uniform $R/X$ ratio}]
The resistance-reactance ratio $\rho = \frac{r_{nk}}{\omega_0\ell_{nk}}$ is the same for all $(n,k) \in \mathcal{E}_{\text{\upshape{tot}}}$.\label{assumption:r/x}
\end{assumption}

With this assumption, we can express $p_{nk}^{\text{pu}}(s)$, 
the per-unit power flowing across line $(n,k)$, as 
\begin{align*}
    p_{nk}^{\text{pu}}(s) = \frac{1}{\ell^{\text{pu}}_{nk}}\mu(s)(\theta_{n}(s) - \theta_k(s)),
\end{align*}
where $\ell^{\text{pu}}_{nk}$ is the per-unit inductance of line $(n,k)$ and $\mu(s)$ is the transfer function given by
\begin{align*}
    \mu(s) \coloneqq \frac{\omega_0^2}{s^2 + 2\omega_0\rho s + \omega_0^2(1 + \rho^2)}.
\end{align*}
Note that this notation represents the power in per-unit notation, but does not rescale the Laplace variable $s$. This allows us to account for networks with a variety of power levels while allowing us to analyze frequency disturbances in their original units. We define the vector $p_{\text{net}}^{\text{tot}}(s) \in \C^{|\mathcal{N}_{\text{tot}}|}$ such that $p_{\text{net,}n}^{\text{tot}}(s)$ represents the per-unit net power injected into the network at bus $n$. The total network power injections are expressed by 
\begin{align*}
    p_{\text{net}}^{\text{tot}}(s) = \mu(s)L_{\text{tot}}\theta_{\text{tot}}(s),
\end{align*}
where $\theta_{\text{tot}}(s) \in \C^{|\mathcal{N}_{\text{tot}}|}$ is the vector of voltage phase angles at each bus and $L_{\text{tot}} \in \R^{|\mathcal{N}_{\text{tot}}| \times |\mathcal{N}_{\text{tot}}|}$ is the Laplacian matrix constructed from the graph $(\mathcal{N}_{\text{tot}},\mathcal{E}_{\text{tot}})$ where each edge $(n,k) \in \mathcal{E}_{\text{tot}}$ has weight $(\ell^{\text{pu}}_{nk})^{-1}$. Note that by defining $\mu(s)$ to be independent of the line inductance, we can account for the resonant effects of line dynamics as a scalar transfer function multiplying the DC power flow approximation. The network is Kron-reduced~\cite{dorfler_kron_2013} to only the nodes $\mathcal{N} \subseteq \mathcal{N}_{\text{tot}}$ corresponding to buses with generation and energy storage resources, where the edges of the reduced network are represented by the set $\mathcal{E} \subseteq \mathcal{N} \times \mathcal{N}$. The power flow for the reduced network can be expressed by 
\begin{align}
    p_{\text{net}}(s) &= \mu(s)L\theta(s)\notag \\
    &= \frac{\mu(s)}{s}L\omega(s),\label{eq:powerflow_vec}
\end{align}
where $p_{\text{net}}(s) \in \C^{|\mathcal{N}|}$, $L \in \R^{\mathcal{N}\times\mathcal{N}}$, $\theta(s) \in \C^{|\mathcal{N}|}$ and $\omega(s) \in \C^{|\mathcal{N}|}$ respectively denote the vector of net power injections, the Laplacian matrix, the vector of voltage phase angles, and the vector of  bus frequencies for the reduced network. 

\subsection{Converter Model}
\begin{figure}
    \centering
    \begin{subfigure}{\linewidth}
        \includegraphics[width=\linewidth]{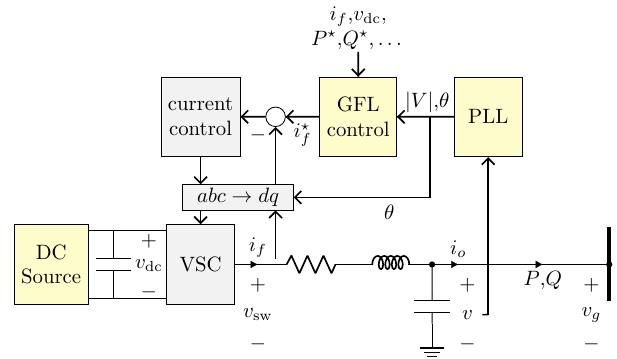}
        \subcaption{GFL control\vspace{1em}}\label{fig:gfl_diag}
    \end{subfigure}
    \begin{subfigure}{\linewidth}
        \includegraphics[width=\linewidth]{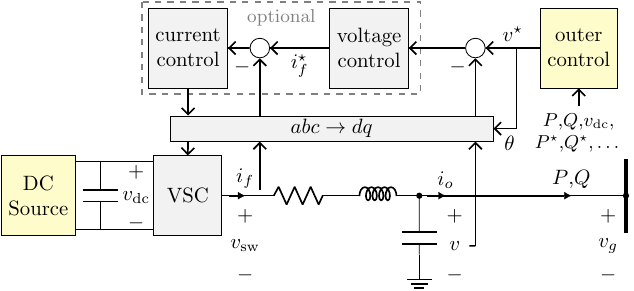}
        \subcaption{GFM control}\label{fig:gfm_diag}
    \end{subfigure}
    
    \caption{Generalized block diagrams for a VSC control loop for both GFL control (\ref{fig:gfl_diag}) and GFM control (\ref{fig:gfm_diag}). The blocks highlighted in yellow show the time-scales of interest in this paper.}\label{fig:conv_block_diag}
\end{figure}
A grid-connected VSC can be modeled with the block diagrams shown in Fig.~\ref{fig:conv_block_diag}. For GFL control, this model includes a phase-locked loop (PLL), a customizable control loop that can implement control objectives (e.g., power reference tracking, DC voltage tracking, etc.) and an inner loop controlling the VSC output current. For GFM control, this model includes an outer control loop (e.g., droop control) that determines the converter's desired AC voltage waveform, and optional current and voltage inner loops. Both models include a DC source with possible DC side dynamics.

Based on this architecture, we can model the dynamics of a voltage source converter as a relationship between the active power output $P$ and the AC frequency of the voltage $v$ imposed by the plant. Note that assigning power as an input and frequency as an output is an arbitrary modeling choice; however, doing so allows us to more easily model the stability of the bus frequencies in response to disturbances in power. As such, we represent grid-connected converter at a bus $n \in \mathcal{N}$ with the transfer function $g_n(s)$. We define $g_n(s)$ such that
\begin{align*}
    \omega_n(s) = g_n(s)p_n(s),
\end{align*}
where $\omega_n(s)$ is the frequency of the AC voltage imposed by the converter and $p_n(s)$ is the per-unit power imbalance at the converter bus, expressed in Laplace domain. Note that this model works for both GFM and GFL control. For GFM control, we can obtain this transfer function directly from the outer controls and DC bus dynamics. For GFL control, we can capture the PLL dynamics as a transfer function from the bus angle to the PLL angle estimate, and capture the internal control dynamics as a transfer function from the angle estimate to the internal power reference~\cite{ducoin_swing_2023}. By inverting these transfer functions, we obtain a total transfer function from power imbalance to frequency for the GFL converter (e.g.,~\cite[Eq. 7b]{bui_input-output_2024}).

The focus of this transfer function model is to capture the outer control loop dynamics and any DC source dynamics under the assumption of timescale separation between these slower dynamics and the faster dynamics created by converter switching and any inner control loops~\cite{subotic_lyapunov_2021}. However, this model can also be extended by augmenting the transfer function to approximate the loss of bandwidth from switching~\cite{qoria_tuning_2018}.

\subsection{Model of Synchronous Machine with Damper Windings}\label{sec:sm_model}

\begin{figure}
    \centering
    \begin{subfigure}{\linewidth}
        \centering
        \includegraphics{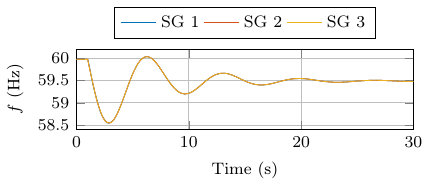}
        \subcaption{EMT frequency response}
    \end{subfigure}
    
    \begin{subfigure}{0.48\linewidth}
        \centering
        \includegraphics{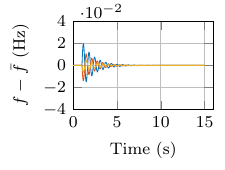}
        \subcaption{EMT simulation}
    \end{subfigure}
    \begin{subfigure}{0.48\linewidth}
        \includegraphics{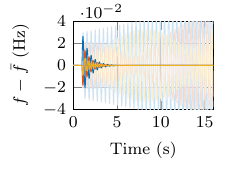}
        \subcaption{Small-signal model}\label{fig:freq_model_dw}
    \end{subfigure}
    \caption{The small-signal system model, using the synchronous generator model~\eqref{eq:sg_tf} and line model~\eqref{eq:powerflow_vec} is compared to an EMT simulation of the IEEE 9-bus model after a load step of $0.5$~p.u. at Bus 7. Incorporating the damper winding model allows the small-signal model to capture the frequency synchronization between the machines (dark lines in (c)), which a swing equation model without damper windings (faint lines in (c)) cannot capture. }\label{fig:ss_dw_validation}
\end{figure}

\begin{subequations}
    The frequency dynamics of a machine at bus $n \in \mathcal{N}$ can be described by 
    \begin{align}
        \frac{2H_n}{\omega_0} \ddt \omega_n &= P_{\text{gen,}n} - P_{\text{load,}n} - P_{\text{net,}n} - P_{\text{D,}n}\label{eq:swing_eq}
    \end{align}
    where $H_n$ denotes the machine's inertia constant, $\omega_n$ denotes the difference between the rotor frequency and the nominal system frequency, $P_{\text{gen,}n}$ denotes the mechanical power generation, $P_{\text{load,}n}$ denotes the power consumed by load at bus $n$, $P_{\text{net,}n}$ denotes the power injected into the network at bus $n$, and $P_{\text{D,}n}$ denotes the damping power induced by the machine damper winding, where all powers are expressed as per-unit quantities. If the machine is a generator with a turbine-governor system, the mechanical power generation follows
    \begin{align}
        T_{\text{G,}n} \ddt P_{\text{gen,}n} &= -P_{\text{gen,}n} - \frac{k_{\text{g,}n}}{\omega_0}\omega_n,\label{eq:turbine_gov}
    \end{align}
    where $T_{\text{G,}n}$ is the turbine time constant and $k_{\text{g,}n}$ is the per-unit governor gain; for a synchronous condenser $P_{\text{gen,}n} = 0$.
\end{subequations}\label{eq:sync_machine}

Under the assumption that $\omega_n$ changes relatively slowly, we can approximate $P_{\text{D,}n} = \xi_{\text{SM,}n} \ddt P_{\text{net,}n}$ 
with the reduced-order damper winding model derived in Sec.~\ref{sec:dw_derivation}, i.e., $\xi_{\text{SM},n}$ given by~\eqref{eq:xi} and the parameters of the machine at bus $n$. As a result, we can express~\eqref{eq:swing_eq} in frequency domain as 
\begin{align*}
     \frac{2H_n}{\omega_0}  \omega_{n}s = P_{\text{gen,}n} - P_{\text{load,}n} - (1 + \xi_{\text{SM},n} s)P_{\text{net,}n}.
\end{align*}
Accounting for the turbine-governor dynamics~\eqref{eq:turbine_gov}, the dynamics of a synchronous generator can be written as 
\begin{align*}
    \omega_n = \frac{\omega_0(1 + \xi_{\text{SM},n} s)(1 + T_{\text{G,}n} s)}{2H_nT_{\text{G,}n} s^2 + 2H_ns + k_{\text{g,}n}} \left(-\frac{P_{\text{load,}n}}{1 + \xi_{\text{SM},n} s} - P_{\text{net,}n}\right).
\end{align*}
We define $P_{\text{dist,}n} \coloneqq P_{\text{load,}n}/(1 + \xi_{\text{SM},n} s)$. The transfer function from $-(P_{\text{dist,}n} + P_{\text{net,}n})$ to frequency for a synchronous generator can then be expressed as
\begin{align}
    g_{\text{SG}}(s) = \frac{\omega_0(1 + \xi_{\text{SM}} s)(1 + T_{\text{G}} s)}{2HT_{\text{G}} s^2 + 2Hs + k_g}.\label{eq:sg_tf}
\end{align}
For synchronous condensers, we derive the transfer function 
\begin{align*}
    g_{\text{SC}}(s) = \frac{\omega_0(1 + \xi_{\text{SM}} s)}{2Hs}.
\end{align*} 

Comparing the small-signal model of the IEEE 9-bus system 
to an EMT simulation shows that adding the reduced-order damper winding model significantly improves the ability of the small-signal machine model to capture the frequency synchronization dynamics (Fig.~\ref{fig:ss_dw_validation}). In fact, if the damper winding model is excluded from the small-signal model, the interactions between the line dynamics and the machine model are unstable.


\subsection{Closed-loop Frequency Domain Model}
\begin{figure}
    \centering
    \includegraphics[width=\linewidth]{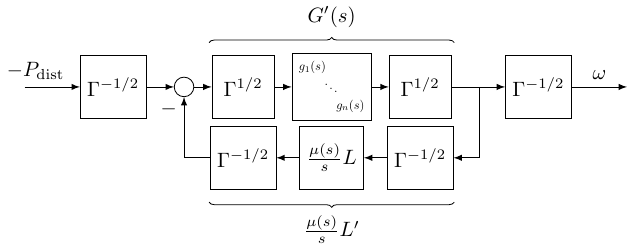}
    \caption{Block diagram of the normalized small-signal power system model. The normalized bus dynamics are represented by $G'(s)$ and the normalized power flow model is represented by $\frac{\mu(s)}{s}L'$.}\label{fig:blockdiag}
\end{figure}

We define the power distubance vector $P_{\text{dist}} \in \C^{|\mathcal{N}|}$ such that $P_{\text{dist,}n} = P_{\text{load,}n}/(1 + \xi_n s)$ if $n \in \mathcal{N}$ is a bus with a synchronous machine and $P_{\text{dist,}n} = P_{\text{load,}n}$ otherwise. The power system dynamics can be described by the negative feedback interconnection between the matrix of bus dynamic transfer functions $G(s) = \diag\left\{g_n(s)\right\}_{n \in \mathcal{N}}$ and the power flow matrix $\frac{\mu(s)}{s}L$, where the input to the system is $P_\text{dist}$ and the output is the frequency $\omega$. If the total transfer function from $P_{\text{load}}$ to $\omega$ is the cascade of two stable systems, it must be stable. Thus, it suffices to check the stability of the transfer function from $P_\text{dist}$ to $\omega$ to show stability of the full system.

To aid in our stability analysis, we make the following assumption about the form of our bus dynamic transfer functions.
\begin{assumption}
    For every bus $n \in \mathcal{N}$, $g_n(s)$ can be expressed as a rational proper transfer function with real coefficients.\label{assumption:rational_prop_tf}
\end{assumption}

Define $\gamma_n \coloneqq 2\sum_{(n,k) \in \mathcal{E}} 1/\ell^{\text{pu}}_{nk}$ and $\Gamma \coloneqq \diag\{\gamma_n\}_{n \in \mathcal{N}}$. Then, we can rescale by $\Gamma$ as shown in Fig.~\ref{fig:blockdiag} to express the system as the feedback interconnection between $G'(s) \coloneqq \Gamma^{1/2}G(s)\Gamma^{1/2}$ and $\mu(s)/s L'$, where $L' \coloneqq \Gamma^{-1/2}L\Gamma^{-1/2}$. Note that $\text{spec}\{L'\} \in [0,1]$~\cite[Lemma 1]{pates_robust_2019} and $G'(s)$ is a diagonal matrix where the $n$-th entry $g_n'(s)$ is equal to the bus dynamics $g_n(s)$ rescaled by the susceptances of local lines.

\section{Stability Analysis}\label{sec:stability}
Section~\ref{sec:stab_criteria} develops conditions to certify the stability of the frequency dynamics in Fig.~\ref{fig:blockdiag}. The conditions are developed in two parts: first, we certify the stability of the system response to slow dynamics, where the bus dynamics cohere to the synchronous system dynamics; next, we certify stability of the faster dynamics where interactions between bus dynamics dominate the system response. This analysis results in a set of stability criteria, which are further decomposed into several timescales in Sec.~\ref{sec:interoperable} to interpret the requirements for different frequency ranges. Section~\ref{sec:application} discusses the application of these conditions.




\subsection{Stability Criteria}\label{sec:stab_criteria}
First, we consider the system response to very slow disturbances (i.e., small $|s|$ in the Laplace domain\footnote{Note that low frequency oscillations are described by $s = j\omega$ in Laplace domain, where $\omega$ is the oscillation frequency, so low frequency perturbations correspond to $s$ with small magnitude.}) and develop conditions to certify its stability. We show that if these conditions are met, the dynamics of each bus cohere very closely to a single synchronous response for these slow disturbances. Following~\cite{min_frequency_2025} we define the synchronous dynamics 
\begin{align}
    \bar{g}(s) \coloneqq {\bigg(\frac{1}{|\mathcal{N}|}\sum\nolimits_{n\in \mathcal{N}} (g_n')^{-1}(s)\bigg)}^{-1}.
\end{align}
Moreover, let $S_{\delta} = \left\{s \in \overline{\C}_+ \: : \: |s| < \delta \right\}$ for some $\delta \in \R_{>0}$, where $\overline{\C}_+$ denotes the closed right-half complex plane, and let $\lambda_2(L')$ denote the second-smallest eigenvalue of $L'$. We can now certify stability of the system response to slow dynamics with the following condition.

\begin{condition}[\textbf{Stable slow coherent dynamics}]\label{cond:stable_steady_state} 
The synchronous dynamics $\bar{g}(s)$ of the system satisfy
\begin{enumerate}
    \item\label{cond:bounded_coherent_dyn} for all $s \in \overline{\C}_+$, $\left|\bar{g}(s)\right| \leq M_1$ for some $M_1 \in \R_{> 0}$.
\end{enumerate}    
Moreover, for some $\delta \in \R_{> 0}$, the following hold for all $s \in S_{\delta}$:
\begin{enumerate}[resume]
    \item\label{cond:no_ss_zero} there exists $M_2 \in \R_{> 0}$ such that $\max\limits_{n \in \mathcal{N}}\left|(g_n')^{-1}(s)\right| \leq M_2$,
    \item\label{cond:coh_region} $\left|\dfrac{s}{\mu(s)}\right| < \dfrac{\lambda_2(L')}{M_2 + M_1 M_2^2}$.
\end{enumerate}
\end{condition}
    \begin{lemma}\label{lemma:coherent}
        If Condition~\ref{cond:stable_steady_state} is satisfied, the closed-loop system in Fig.~\ref{fig:blockdiag} will have a bounded response to the slow disturbances $P_{\text{dist}}(s)$ characterized by $s \in S_{\delta}$. Moreover, the response of each bus can be represented as a finite magnitude deviation from the synchronous response $\bar{g}(s)$. 
    \end{lemma}
A formal proof of this lemma is provided in the appendix, but we provide an informal interpretation here. Condition~\ref{cond:stable_steady_state}.\ref{cond:bounded_coherent_dyn} ensures that the synchronous dynamics $\bar{g}(s)$ are stable. Moreover, by~\cite[Lemma 1]{min_frequency_2025}, Conditions~\ref{cond:stable_steady_state}.\ref{cond:no_ss_zero} and \ref{cond:stable_steady_state}.\ref{cond:coh_region} ensure that, for slow dynamics (i.e., $s \in S_{\delta}$), the system response will be incrementally stable with respect to the synchronous dynamics (i.e., the deviation of the system response from the synchronous response is bounded), which can be interpreted as the system response cohering to the synchronous dynamics near steady-state. Therefore, Condition~\ref{cond:stable_steady_state} ensures that the system will have a stable response to slow perturbations within the region $S_\delta$ in the Laplace domain. Note that if Conditions~\ref{cond:stable_steady_state}.\ref{cond:bounded_coherent_dyn} and~\ref{cond:stable_steady_state}.\ref{cond:no_ss_zero} are satisfied for some $S_{\delta^\prime}$, there always exists a small enough $\delta \leq \delta^\prime$ such that Condition~\ref{cond:stable_steady_state}.\ref{cond:coh_region} is satisfied over $S_{\delta}$. In practice, system operators today already assume that bulk power systems satisfy this slow coherency requirement. Secondary frequency control (i.e., AGC) assumes that the bus frequencies synchronize on timescales up to approximately 30 seconds, which implicitly requires slow coherency for frequencies up to roughly 0.03~Hz (i.e., Condition~\ref{cond:stable_steady_state} is satisfied for $\delta$ of 0.03~Hz). This slow coherence assumption also allows us to explain and establish the frequency stability of systems including STATCOMs, synchronous condensers, and other marginally stable units, which individually cannot stabilize steady-state disturbances but cohere to the system steady-state response in an interconnected network.

Next, we present interoperability conditions on each bus to ensure that the system response to faster disturbances (i.e., $s \in \overline{\C}_+ \setminus S_{\delta}$) is stable.
\begin{condition}[\textbf{Stable synchronizing dynamics}]\label{cond:formal_transient_stability} For a given $\delta \in \R_{> 0}$, the following conditions hold for all $n \in \mathcal{N}$:
\begin{enumerate}
    \item\label{cond:no_transient_poles} $g_n'(s)$ does not have any poles in $\overline{\C}_+ \setminus S_{\delta}$,
    \item\label{cond:interoperability} there exists a function $\alpha\: : \: \C \to (0, \pi/2]$ such that 
    \begin{align*}
        \Re\left\{e^{j(\pi/2 - \alpha(s))}\left(1 + \frac{\mu(s)}{s}g'_n(s)\right)\right\} > 0
    \end{align*}
    for all $n \in \mathcal{N}$ and all $s$ on the boundary of $\overline{\C}_+ \setminus S_{\delta}$.
\end{enumerate}
\end{condition}

    \begin{lemma}\label{lemma:interoperable}
        If Condition~\ref{cond:interoperability} is satisfied, the closed-loop system in Fig.~\ref{fig:blockdiag} will have a bounded response to all disturbances $P_{\text{dist}}(s)$ where $s \in \overline{\C}_+ \setminus S_{\delta}$.
    \end{lemma}

A proof of this lemma is provided in the appendix, but a broad interpretation of Condition~\ref{cond:formal_transient_stability} is presented here. Condition~\ref{cond:formal_transient_stability}.\ref{cond:no_transient_poles} ensures that each $g_n(s)$ does not have any unstable poles outside of the coherent region $S_{\delta}$. Condition~\ref{cond:formal_transient_stability}.\ref{cond:interoperability} ensures the interoperability of the bus dynamics by ensuring that for a given $s$, all $\frac{\mu(s)}{s}g'_n(s)$ must lie in a fixed half plane given by $\alpha(s)$. A more interpretable sufficient condition for Condition~\ref{cond:formal_transient_stability}.\ref{cond:interoperability} is presented in Sec.~\ref{sec:interoperable}. Combining Lemma~\ref{lemma:coherent} and Lemma~\ref{lemma:interoperable}, can now present our main stability result.

\begin{theorem}[\textbf{Frequency stability}]\label{thm:main_result}
    Consider a connected Kron-reduced network with uniform $R/X$ ratio $\rho > 0$. 
    If there exists $\delta > 0$ such that Conditions~\ref{cond:stable_steady_state} and~\ref{cond:formal_transient_stability} are satisfied, the interconnected system in Fig.~\ref{fig:blockdiag} is internally stable.
\end{theorem}

    \begin{proof}
        By~\cite[Thm 4.12]{skogestad_multivariable_2005}, showing that the transfer function 
            $H(s) \coloneqq {\left(I + G'(s)\frac{\mu(s)}{s}L'\right)}^{-1}G'(s)$
        is stable guarantees the internal stability of the closed loop system in Fig.~\ref{fig:blockdiag} if no poles in $\overline{\C}_+$ are canceled between $G'(s)$ and $\frac{\mu(s)}{s}L'$. This requirement is satisfied by Conditions~\ref{cond:stable_steady_state}.\ref{cond:bounded_coherent_dyn} and~\ref{cond:stable_steady_state}.\ref{cond:no_ss_zero}: Condition~\ref{cond:stable_steady_state}.\ref{cond:bounded_coherent_dyn} ensures the synchronous dynamics have no unstable poles so that the kernel of $L'$, given by $\mathrm{span}{(\mathds{1}_n)}$, cannot cancel out any unstable poles; Condition~\ref{cond:stable_steady_state}.\ref{cond:no_ss_zero} implies that $G'(s)$ has no zeros at the origin, so it will not cancel the $\frac{1}{s}$ term in $\frac{\mu(s)}{s}L'$. Moreover, because $\rho > 0$, all poles of $\mu(s)$ are stable, so no unstable poles are canceled.
        To prove the stability of $H(s)$, we show that $|H(s)|$ is bounded over $s \in \overline{\C}_+$, which is satisfied by Lemma~\ref{lemma:coherent} for $s \in S_{\delta}$ and by Lemma~\ref{lemma:interoperable} for $s \in \overline{\C}_+ \setminus S_{\delta}$.
    \end{proof}

\subsection{Understanding Interoperability Condition}\label{sec:interoperable}
In this section we introduce and analyze Condition~\ref{cond:interpretable}, a more interpretable sufficient condition for satisfying Condition~\ref{cond:formal_transient_stability}.\ref{cond:interoperability}.

\begin{condition}[\textbf{Interpretable interoperability condition}]\label{cond:interpretable} For some $\delta \in \R_{> 0}$, some $\omega_1, \omega_2 \in \R_{> 0}$ where $\delta \leq \omega_1 \leq \omega_2$, and some function $\alpha\: : \: [\omega_1,\, \omega_2) \to (0, \pi/2]$, the following conditions hold for all $n \in \mathcal{N}$:
    \begin{enumerate}[leftmargin=1em]
        \item[] \textbf{Region 1:} for all $s \in \left\{\delta e^{j\theta} \: : \: \theta \in [0, \pi/2)\right\}\cup\left\{j\omega \: : \: \omega \in [\delta, \omega_1)\right\}$, it holds that $\Re(\mu(s)g_n'(s)) > 0$ 
        
        \item[] \textbf{Region 2:} for all $\omega \in [\omega_1,\, \omega_2)$, one of the following conditions holds true: 
        \begin{itemize}[leftmargin=2em]
            \item $\angle \mu(j\omega)g_n'(j\omega) \in (-\pi/2 + \alpha(\omega),\, \pi/2 - \alpha(\omega))$ 
            \item $\left|\frac{\mu(j\omega)}{\omega}g_n'(j\omega)\right| < \frac{\sin \alpha(\omega)}{\left|\cos \left(\angle \mu(j\omega)g_n'(j\omega) - \alpha(\omega)\right) \right|}$
        \end{itemize}
        
        \item[] \textbf{Region 3:} for all $\omega \in [\omega_2,\, \infty]$,  $\left|\frac{\mu(j\omega)}{\omega}g_n'(j\omega)\right| < 1$ holds.
    \end{enumerate}
\end{condition}

\begin{figure}
    \centering
    \includegraphics[width=0.7\columnwidth]{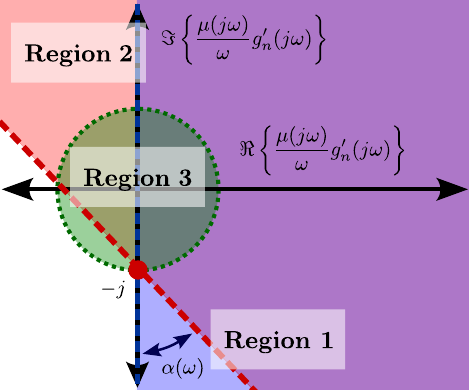}
    \caption{The interoperability regions in Condition~\ref{cond:interpretable} are shown on the complex plane. Region 1 is shaded in blue, Region 2 in red, and Region 3 in green.}
    \label{fig:revised_stab_condition}
\end{figure}

These regions, depicted in Fig.~\ref{fig:revised_stab_condition}, can be understood as requiring small phase (Region 1), small gain (Region 3), or some tradeoff between the two (Region 2). We first consider Regions 1 and 3, which can be interpreted as letting $\alpha(s)$ in Condition~\ref{cond:formal_transient_stability}.\ref{cond:interoperability} approach $0$ and $\pi/2$, respectively.\footnote{Note that $\alpha$ still remains strictly greater than zero in Region 1, satisfying $\alpha \in (0, \pi/2]$. Condition~\ref{cond:formal_transient_stability}.\ref{cond:no_transient_poles} requires $g_n'(s)$ to have finite gain, ensuring that if $\Re(\mu(s)g_n'(s)) > 0$, we can always select $\alpha > 0$ to satisfy Condition~\ref{cond:formal_transient_stability}.\ref{cond:interoperability}.}  Region 1 requires each $g_n(s)\mu(s)$ to have positive real part for low frequencies, which means each $g_n(s)\mu(s)$ must dissipate energy at these frequencies. It also requires $g_n(s)\mu(s)$ to have positive real part on the set $s \in \left\{\delta e^{j\theta} \: : \: \theta \in [0, \pi/2)\right\}$, but Assumption~\ref{assumption:rational_prop_tf} ensures this can always be satisfied for small enough $\delta$ if  $\Re(g_n(j\delta) \mu(j\delta)) > 0$. Region 3 requires each $g_n(s)\mu(s)$ to have small gain at high frequencies, where a more connected node (i.e., large $\gamma_n$) has a stricter gain limit. We can interpret the Region 1 requirements as specifying a transient droop behavior and the Region 3 requirements as dictating the inertial response.


Region 2, the medium frequency region, allows $\alpha(s)$ to take a value between those extremes, trading off the phase and gain conditions on $g_n'(s)\mu(s)$. If the phase satisfies $\angle \mu(j\omega)g_n'(j\omega) \in (-\pi/2 + \alpha(\omega),\, \pi/2 - \alpha(\omega))$, $g_n'(s)$ will satisfy the interoperability requirements for mid-range frequencies. If $g_n'(s)$ does not satisfy this phase requirement, it can 
satisfy the second condition of Region 2, which imposes a gain limit that grows tighter as the phase of $\mu(j\omega)g_n'(j\omega)$ grows further outside of the range $(-\pi/2 + \alpha(\omega),\, \pi/2 - \alpha(\omega))$. If the gain satisfies $\left|\frac{\mu(j\omega)}{\omega}g_n'(j\omega)\right| < \sin \alpha(\omega)$, the requirement will be satisfied for any phase of $\mu(j\omega)g_n'(j\omega)$. Note that if $g_n'(s)$ satisfies $\Re(\mu(j\omega)g_n'(j\omega)) > 0$ and $\left|\frac{\mu(j\omega)}{\omega}g_n'(j\omega)\right| < 1$ (i.e., satisfies the conditions from both Regions 1 and 3), the condition from Region 2 will always be satisfied.  

\subsection{Application of Stability Criteria}\label{sec:application}
In this section, we discuss how Conditions~\ref{cond:stable_steady_state} and \ref{cond:interpretable} can be applied to certify stability of a power system. As discussed above, if Conditions~\ref{cond:stable_steady_state}.\ref{cond:bounded_coherent_dyn} and \ref{cond:stable_steady_state}.\ref{cond:no_ss_zero} hold for any $S_{\delta'}$, there always exists some small enough $\delta > 0$ such that Condition~\ref{cond:stable_steady_state}.\ref{cond:coh_region} holds over $S_{\delta}$. Thus, in practice we can verify that Conditions~\ref{cond:stable_steady_state}.\ref{cond:bounded_coherent_dyn} and~\ref{cond:stable_steady_state}.\ref{cond:no_ss_zero} hold for some small $\delta' > 0$, and conclude that Condition~\ref{cond:stable_steady_state} certifies stability of the system to steady state disturbances (i.e., $s = 0$). Then, by ensuring that each $g_n(s)$ do not have any poles in $\overline{\C}_+ \setminus 0$ (satisfying Condition~\ref{cond:interoperability}.\ref{cond:no_transient_poles}) and that Condition~\ref{cond:interpretable} holds for any $\delta \in (0,\, \epsilon]$ for some $\epsilon > 0$, we can certify stability of the system.

We now discuss how Condition~\ref{cond:interpretable} can be used for stability analysis, defining a topology-independent relative stability margin to aid in its application. Condition~\ref{cond:interpretable} requires that for every bus, the function $g_n(j\omega)\mu(j\omega)$ must, at minimum, have either positive real part (i.e., dissipate energy) or gain less than $\omega/\gamma_n$ for each frequency $\omega \neq 0$. Generally, the $180^{\circ}$ phase shift induced by the line dynamics will cause $\Re(g_n(j\omega)\mu(j\omega))$ to become negative at higher frequencies for many common bus models. As a result, for a bus $n$ with dynamics $g_n(s)$, we define the crossover frequency $\omega_{\text{c}}^n \coloneqq \min \{\omega \, : \, \omega \in \R_{\geq 0},\, \angle g_n(j\omega)\mu(j\omega) = -90^{\circ}\}$.
In a system with homogeneous bus dynamics, $\omega_\text{c}^n$ will be the same for all $n \in \mathcal{N}$, and by setting $\omega_1 = \omega_2 = \omega_\text{c}$, Condition~\ref{cond:interpretable} reduces to a condition that each $g_n'(s)$ satisfies $\left|\frac{1}{\omega}\mu(j\omega)g_n'(j\omega)\right| < 1$ for all $\omega \geq \omega_\text{c}$. For a heterogeneous system, we define $\omega_1 \leq \min_{n \in \mathcal{N}} \omega_\text{c}^n$ and $\omega_2 \geq \max_{n \in \mathcal{N}} \omega_\text{c}^n$. 
Then, in Region 2, $\alpha(\omega)$ can be set to trade off between the strictness of the phase and the gain conditions: a larger $\alpha(\omega)$ corresponds to a less restrictive gain condition at the expense of a tighter phase bound for buses that do not meet the gain condition, and vice versa. However, for any choice of $\alpha(\omega)$, every bus $n \in \mathcal{N}$ must satisfy $\left|\frac{\mu(j\omega_\text{c}^n)}{\omega_\text{c}^n}g_n'(j\omega_\text{c}^n)\right| < 1$. We can therefore define 
\begin{align*}
M (g,\mu) \coloneqq \frac{\omega_\text{c}}{\left|\mu(j\omega_\text{c})g(j\omega_\text{c})\right|},
\end{align*}
as the relative stability margin, where $\left|\frac{\mu(j\omega_\text{c}^n)}{\omega_\text{c}^n}g_n'(j\omega_\text{c}^n)\right| < 1$ will be satisfied if $M(g_n,\mu) > \gamma_n$. 
Figure~\ref{fig:gain_margin_sm_dw} shows a sample Bode plot of $\frac{1}{\omega}\mu(j\omega)g(j\omega)$ for a synchronous machine where the crossover frequency $\omega_\text{c}$ and the relative stability margin $M(g_{\text{SG}}, \mu)$ are visualized. 

In practice, system operators could determine $\omega_1$, $\omega_2$, and $\alpha(\omega)$ based on the existing network dynamics and the desired system behavior. Then, these parameters could be used as performance specifications for new generation seeking to connect to the grid and be applied to each new device in a decentralized manner. System operators could evaluate the performance of new devices by using a frequency sweep to generate a Bode plot~\cite{bui_input-output_2024}, and with that information determine whether the device could meet the performance specifications laid out in Condition~\ref{cond:interpretable}, and if so, determine how the device would need to be located relative to existing devices to ensure stability. This can be done by computing the maximum $\gamma_n$ (i.e., local connectivity) that allows the device to satisfy the stability conditions, which corresponds to the minimum electrical distance between the device and other adjacent buses. Even Condition~\ref{cond:stable_steady_state}.\ref{cond:bounded_coherent_dyn}, which relies on knowledge of the dynamics of every bus, can be evaluated in a decentralized manner by storing only the current $\bar{g}(s)$ and iteratively updating it with each new device that is added to the system.

\section{Effect of Damper Winding Model on Synchronous Machine Stability}\label{sec:analyze_dw_effects}
This section examines the application of Condition~\ref{cond:interpretable} to synchronous machine models in a system with line dynamics. Notably, including the reduced-order damper winding model can improve the stability of synchronous machine models even in cases when increasing machine inertia hurts stability.

\subsection{Synchronous Generators}
Using the framework from Sec.~\ref{sec:sm_model}, the bus dynamics of synchronous generators are modeled by~\eqref{eq:sg_tf}.
For small enough $\delta$, $\mu(\delta e^{j\theta})g_{\text{SG}}(\delta e^{j\theta})$ will have positive real part for any $\theta \in [-\pi/2, \pi/2]$.
Moreover, $\mu(j\omega)g_{\text{SG}}(j\omega)$ will have positive real part for small $\omega$. However, the phase shift induced by the line dynamics will cause $\mu(j\omega)g_{\text{SG}}(j\omega)$ to have negative real part at higher frequencies. For a synchronous generator at bus $n$, Condition~\ref{cond:interpretable} requires that  $M(g_{\text{SG}}, \mu) > \gamma_n$ and that for all $\omega > \omega^{\text{SG}}_\text{c}$, $\frac{1}{\omega}\left|\mu(j\omega)g_{\text{SG}}(j\omega)\right| < \gamma_n^{-1}$ holds, where $\omega^{\text{SG}}_\text{c}$ is the crossover frequency of the bus dynamics. Therefore, if $\omega^{\text{SG}}_\text{c}$ is less than the line dynamic resonant frequency $\omega_\text{r} \coloneqq \omega_0\sqrt{1-\rho^2}$, a potential local maximum of the transfer function, we must also verify that $\frac{1}{\omega_\text{r}}\left|\mu(j\omega_\text{r})g_{\text{SG}}(j\omega_\text{r})\right| < \gamma_n^{-1}$.

\begin{figure}
    \centering
    \includegraphics{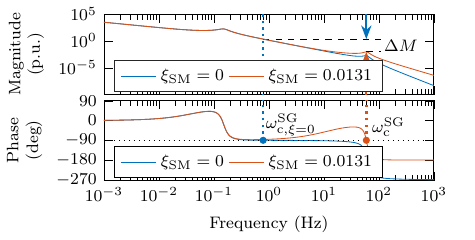}
    \caption{The Bode plot of $\frac{1}{\omega}\mu(j\omega)g(j\omega)$ is shown for a 60 Hz system where $\rho=0.1$ and $g(j\omega) = g_{\text{SG}}(j\omega)$ with $H=3.7$, $T_{\text{G}} = 3$, and $k_g = 20$. The transfer function is plotted with no damper winding model (i.e. $\xi_{\text{SM}} = 0$) and with $\xi_{\text{SM}} = 0.0131$. The crossover frequency $\omega^{\text{SG}}_{\text{c}}$ for each value of $\xi_{\text{SM}}$ is indicated on the phase plot, and $\Delta M$ on the magnitude plot depicts the increased stability margin for $\xi_{\text{SM}} = 0.0131$. 
    \label{fig:gain_margin_sm_dw}}
\end{figure}

\begin{figure}
    \centering
    \includegraphics{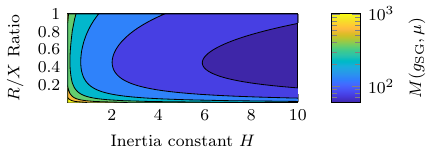}
    \includegraphics{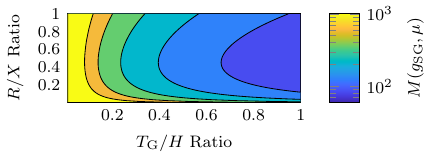}
    \caption{The relative stability margin of a synchronous machine is shown over a range of machine parameters and system $R/X$ ratios. Above, the margin is shown as a function of the inertia constant $H$ while the ratio $T_{\text{G}}/H$ of the turbine time constant to the inertia constant is fixed at 0.8. Below, the margin is shown as a function of $T_{\text{G}}/H$ while for $H$ fixed at $3s$. 
    }\label{fig:sm_stab_margin_inertia}
\end{figure}

By including the reduced-order damper winding model in $g_{\text{SG}}(s)$ (i.e., $\xi_{\text{SM}} > 0$), we can increase the relative stability margin. The damper winding model adds an additional zero to $g_{\text{SG}}(s)$, lifting the phase by $90^\circ$. If $\xi_{\text{SM}} > (\omega_0\sqrt{1 + \rho^2})^{-1}$, $\omega_\text{c}^{\text{SG}}$ will move towards $\omega_0\sqrt{1 + \rho^2}$, the natural frequency of the line dynamics. To understand the impact of the reduced-order damper winding model, we will compare the relative stability margin of $g^{\xi = 0}_{\text{SG}}(s)$, the synchronous generator transfer function without the damper winding model, to the margin of $g_{\text{SG}}(s)$ with the damper winding model (i.e., $\xi_{\text{SM}} > 0$). 

First, we approximate $M(g^{\xi = 0}_{\text{SG}}, \mu)$. Let $\omega^{\text{SG}}_{\text{c,}\xi = 0}$ denote the crossover frequency of the model when the damper windings are excluded. Note that while $\omega^{\text{SG}}_{\text{c,}\xi = 0}$ cannot easily be expressed in closed form, it can be numerically computed for a given bus and will always lie between the poles of $g_{\text{SG}}(s)$ and $\mu(s)$.
By invoking timescale separation between the synchronous machine dynamics and the line dynamics, we assume that $\omega^{\text{SG}}_{\text{c,}\xi = 0} \gg \sqrt{k_{\text{g}}/2HT_{\text{G}}}$, allowing us to approximate $|g_{\text{SG}}(j\omega^{\text{SG}}_{\text{c,}\xi = 0})| \approx \omega_0/(2H\omega^{\text{SG}}_{\text{c,}\xi = 0})$. Similarly, we assume $\omega^{\text{SG}}_{\text{c,}\xi = 0} \ll \omega_0\sqrt{1 + \rho^2}$ to obtain $|\mu(j\omega_{\text{c,}\xi = 0})| \approx 1/\left(1 + \rho^2\right)$. Therefore, we can approximate the stability margin as
\begin{align}
    M(g^{\xi = 0}_{\text{SG}}, \mu) \approx \frac{2H\left(1 + \rho^2\right)(\omega^{\text{SG}}_{\text{c,}\xi = 0})^2}{\omega_0}.\label{eq:sm_stab_margin_no_dw}
\end{align}

\begin{remark}
    Increasing the inertia constant $H$ moves the poles of  $g_\text{SG}(s)$ to lower frequencies, thereby decreasing $\omega^{\text{SG}}_{\text{\upshape{c,}}\xi = 0}$. As a result, despite $H$ appearing in the numerator of~\eqref{eq:sm_stab_margin_no_dw}, increasing the inertia may lower $M(g^{\xi = 0}_{\text{SG}}, \mu)$ (see Fig.~\ref{fig:sm_stab_margin_inertia}). 
\end{remark}

To conservatively estimate the effect of the damper winding model, we compare $M(g^{\xi = 0}_{\text{SG}}, \mu)$ to $M(g_{\text{SG}}, \mu)$ if $\xi_{\text{SM}}$ moves $\omega_{\text{c}}^{\text{SG}}$ to $\omega_\text{r} = \omega_0\sqrt{1 - \rho^2}$. Invoking timescale separation again, we approximate $g(j\omega^{\text{SG}}_{\text{c}}) \approx \xi_{\text{SM}}\omega_0/(2H)$ and $\mu(j\omega^{\text{SG}}_{\text{c}}) = 1/2\rho$, resulting in the conservative estimate
\begin{align}
    M(g_{\text{SG}}, \mu) \approx \frac{4H\rho  \sqrt{1 - \rho^2}}{\xi_{\text{SM}} }.\label{eq:sm_stab_margin_dw}
\end{align}

Comparing~\eqref{eq:sm_stab_margin_dw} to~\eqref{eq:sm_stab_margin_no_dw}, we estimate that incorporating the damper winding model will improve the stability margin if 
\begin{align}
        (\omega_0\sqrt{1 + \rho^2})^{-1} < \xi_{\text{SM}} < \frac{2\rho\omega_0\sqrt{1-\rho^2}}{(\omega^{\text{SG}}_{\text{c,}\xi = 0})^2(1 + \rho^2)}.\label{eq:sm_gamma_bound}
\end{align} 
This range aligns with typical values of $\xi_{\text{SM}}$ for transmission-connected generators. An example is shown in Fig.~\ref{fig:gain_margin_sm_dw}.

Note that the upper bound depends heavily on the network $R/X$ ratio: if $\rho$ approaches zero, the damper winding model will not improve the stability margin of the system. Thus, the damper windings can be understood to help compensate for the phase lag induced by line losses. The bound is also inversely proportional to $(\omega^{\text{SG}}_{\text{c,}\xi = 0})^2$, suggesting that damper windings have greater stability benefits for machines with more inertia.

\subsection{Synchronous Condensers}
If $\xi_{\text{SM}} = 0$, $g_{\text{SC}}(j\omega)$ will be purely imaginary for all $\omega$. Because $\mu(j\omega)$ induces a negative phase shift, this means that $\Re\left(\mu(j\omega)g_{\text{SC}}(j\omega)\right) \leq 0$. Furthermore, for small $\omega$, the gain of $g_{\text{SC}}(j\omega)$ becomes very large, so unless the network is very weakly connected (i.e., $\gamma$ is very small), the synchronous condenser will not satisfy either the phase or the gain condition at low frequencies if $\xi_{\text{SM}} = 0$. 
However, $\xi_{\text{SM}} > 0$ will induce a positive phase shift that, if $\xi_{\text{SM}}$ is large enough, makes $\Re\left(\mu(j\omega)g_{\text{SC}}(j\omega)\right) > 0$ for low frequencies where $g_{\text{SC}}(j\omega)$ does not meet the small gain requirement. Therefore, modeling the effect of damper windings is necessary for demonstrating the stability of synchronous condensers in this framework. 

\section{Emulating Damper Windings through Converter Control}\label{sec:dw_emulation}
The stabilizing effects of damper windings in machines can be achieved in converter-interfaced resources by incorporating a PD term to emulate the reduced-order damper winding model. This can provide significant stability improvements over more conventional droop/virtual synchronous machine (VSM) control, particularly in networks with a larger $R/X$ ratio. While the resulting PD droop control has been proposed before, its impact on stability at a system-wide level has has only been described in a system with homogeneous bus dynamics~\cite{gross_compensating_2022}, and it's unclear how to tune the relevant parameters in a heterogeneous system with a variety of machines and converters. This section generalizes the results of~\cite{gross_compensating_2022} to a network with heterogeneous bus dynamics, demonstrating how PD droop control can achieve the same stabilizing effect as damper windings. Moreover, by applying the analytical stability criteria from Sec.~\ref{sec:stability}, we can analytically tune the control parameters, rather than simply using trial and error.

This section also presents a physical interpretation of PD droop control as emulating the effects of damper windings for converters. Compared with other damper winding emulation strategies~\cite{cvetkovic_modeling_2015,ebrahimi_improved_2019,yin_attenuation_2022, gajare_grid-forming_2025}, PD droop control is significantly simpler, allowing us to analytically model the effect of the damper winding emulation in a system with heterogeneous generation. Emulating the reduced-order damper winding model rather than the full model also has performance advantages. Because damper windings are essentially induction machines, they may destabilize the frequency dynamics if the slip between rotor and stator frequencies becomes too large. In contrast, the reduced-order PD model only models the stabilizing linear region of the damper windings, i.e., effectively emulates only the stabilizing effects of the damper windings. This section begins by analyzing the ability of inertia emulation to stabilize conventional droop control, and then compares the stability of conventional droop control with PD droop control. We then briefly discuss the impacts of PD droop on the DC voltage dynamics and conclude with an explanation of how these results can be used for tuning PD controls.

\subsection{Conventional Droop Control}
Conventional filtered droop control can be expressed as 
\begin{align*}
    g_\text{dr}(s) = \frac{m_p\omega_0}{T_p s + 1},
\end{align*}
where $m_p$ is the per-unit droop coefficient and $T_p$ is the filter time constant. Note that this is equivalent to VSM control: $g_\text{dr}(s)$ matches~\eqref{eq:sg_tf} if $T_{\text{G}}=0$, $\xi_{\text{SM}}=0$, and $T_p = 2H/k_\text{g}$.

Let $\omega_\text{c}^{\text{dr}}$ denote the frequency at which $\Re\left(\mu(j\omega)g_{\text{dr}}(j\omega)\right)$ crosses from positive to negative. We can express $\omega_\text{c}^{\text{dr}}$ as a function of $T_p$ by solving $\Re\left(\mu(j\omega)g_{\text{dr}}(j\omega)\right) < 0$ for $\omega$, which we can equivalently find by solving for the $\omega$ where $\Re\left(\mu^{-1}(j\omega)g_{\text{dr}}^{-1}(j\omega)\right) < 0$. This expression becomes 
\begin{align*}
    0 &> \Re(\mu^{-1}(j\omega)g_{\text{dr}}^{-1}(j\omega)) \\
    &> -\omega^2 - 2T_p\rho\omega_0 \omega^2 + \omega_0^2(1 + \rho^2),
\end{align*}
and we find that $\Re\left(\mu(j\omega)g_{\text{dr}}(j\omega)\right)$ becomes negative at 
\begin{align*}
    \omega_\text{c}^{\text{dr}} = \omega_0\sqrt{\frac{1 + \rho^2}{1 + 2\rho \omega_0 T_p}}.
\end{align*}

With this expression for $\omega_\text{c}^{\text{dr}}$, the condition $M(g_{\text{dr}}, \mu) > \gamma_n$ is equivalent to the stability condition for a network of homogeneous droop-controlled VSCs~\cite[Eq. (10)]{gross_compensating_2022}. The stability criteria derived here therefore replicate the results from~\cite{gross_compensating_2022} concerning the potential instability of droop control: namely, that stabilizing droop control by tuning the virtual inertia may not always be possible, or might require infeasibly large values of $T_p$ (as large values of $T_p$ require large power buffers that may be expensive or impractical).
To ensure these dynamics also meet our interoperability requirements, we must also ensure that any gain increase caused by the resonant peak does not result in instability: i.e., if $\omega^{\text{dr}}_\text{c} < \omega_\text{r}$, it must also hold that $\frac{1}{\omega_\text{r}}\left|\mu(j\omega_\text{r})g_{\text{dr}}(j\omega_\text{r})\right| < \gamma_n^{-1}$. Rewriting this condition, we can equivalently conclude that if $T_p > \rho/(\omega_0 (1 - \rho^2))$, then it must hold that 
\begin{align}
    T_p > \frac{\sqrt{\gamma_n^2 m_p^2 - 4\rho^2(1-\rho^2)}}{2\rho(1-\rho^2)\omega_0}.\label{eq:T_lower_bound}
\end{align}
However, increasing $T_p$ beyond this bound may not necessarily result in better performance, as increasing $T_p$ can reduce $\omega_\text{c}^{\text{dr}}$ to a frequency with a higher gain.

\subsection{PD Droop Control as Damper Winding Emulation}
To improve the stability of droop control, we augment $g_{\text{dr}}(s)$ to mimic the reduced-order damper winding model with the PD term $(1 + \xi_{\text{C}} s)$ for $\xi_{\text{C}} \in \R_{>0}$, resulting in PD droop control
\begin{align*}
    g_\text{pd}(s) = \frac{m_p\omega_0(1 + \xi_{\text{C}} s)}{T_p s + 1}.
\end{align*}
Note that while $g_\text{pd}(s)$ is not realizable for $T_p=0$, it can still be implemented in a VSC because the transfer function for the angle, given by $\frac{1}{s} g_\text{pd}(s)$, is realizable. Notably, dual-port GFM control~\cite{subotic_universal_2024} with a curtailed resource matches the form of $g_{\text{pd}}(s)$~\cite{gross_compensating_2022}. As with synchronous machines, $\xi_{\text{C}}$ provides a phase lift that increases $\omega_\text{c}$ at the expense of increasing $|g(s)|$. In particular, if $\xi_{\text{C}} > (\omega_0\sqrt{1 + \rho^2})^{-1}$, its phase lift will partially compensate the phase drop from $\mu(s)$, increasing $\omega_\text{c}$. 

We can estimate the range of parameters for which adding $\xi_{\text{C}} > 0$ improves the performance of droop control. VSCs have faster dynamics than machines, so we can no longer assume timescale separation with the line dynamics. However, if 
\begin{align}
    M(g_{\text{dr}}, \mu) < \frac{\omega}{\left|\mu(j\omega)g_{\text{dr}}(j\omega)\left(1 + \frac{j\omega}{\omega_0\sqrt{1 + \rho^2}}\right)\right|} \label{eq:droop_pd_potential}
\end{align}
holds for all $\omega > \omega_\text{c}^{\text{dr}}$, increasing $\omega_\text{c}$ by setting $\xi_{\text{C}} \geq (\omega_0\sqrt{1 + \rho^2})^{-1}$ can increase the relative stability margin, despite it also increasing $|g(s)|$.  Note that the natural frequency of $\mu(s)$ is given by $\omega_\text{n} = \omega_0\sqrt{1+\rho^2}$. 
To approximate the parameters for which~\eqref{eq:droop_pd_potential} holds for $\omega > \omega_\text{c}^{\text{dr}}$, we evaluate the inequality at the potential local maximum $\omega_\text{r}$. This reduces~\eqref{eq:droop_pd_potential} to the condition that
\begin{align*}
    \Xi \coloneqq (1 + 2\rho\omega_0T_p)^2 - \sqrt{\frac{2 \omega_\text{n}^2 \left(T_p^2\omega_\text{n}^2 + 2\rho\omega_0T_p + 1\right)^2}{\omega_\text{r}^2(T_p^2\omega_{\text{r}}^2+1)}} \label{eq:droop_inertia_bound}
\end{align*}
\begin{figure}
    \centering
    \includegraphics{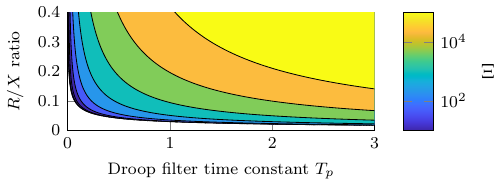}
    \caption{$\Xi$ is plotted over various $R/X$ ratios and virtual inertia values, with the graph left blank if the value is negative.}\label{fig:potential_improvement}
\end{figure}%
\begin{figure}
    \centering
    \includegraphics{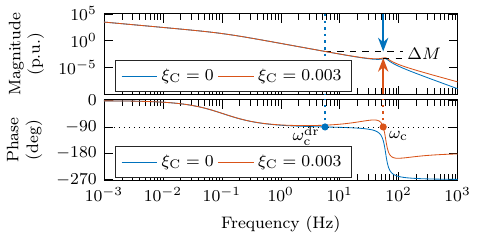}
    \caption{The Bode plot of $\frac{1}{\omega}\mu(j\omega)g(j\omega)$ is shown for a 60 Hz system where $\rho=0.1$ and $g(j\omega) = g_{\text{pd}}(j\omega)$ with $T_p=1.5$.
     The crossover frequency $\omega_{\text{c}}$ for each value of $\xi_{\text{C}}$ is indicated on the phase plot, and $\Delta M$ on the magnitude plot depicts the increased stability margin for $\xi_{\text{C}} = 0.003$. 
    }\label{fig:gain_margin_vsc_dw}
\end{figure}%
must be positive. A larger $\Xi$ indicates that $\xi_{\text{C}}$ can increase the stability margin further. Figure~\ref{fig:potential_improvement} plots $\Xi$ over a range of $\rho$ and $T_p$, showing that adding a non-zero derivative term $\xi_{\text{C}}$ can achieve the most substantial stability gains for devices with larger inertia ratios in more resistive networks. However, $\Xi$ is still positive except for very small $\rho$ or $T_p$, showing that incorporating $\xi_{\text{C}} > 0$ can improve the stability margin for VSCs in most typical operating scenarios (e.g., Fig.~\ref{fig:gain_margin_vsc_dw}). Moreover, unlike increasing the virtual inertia, adding a derivative term does not require increasing the DC energy or power requirements of the converter.

\subsection{Impact of Inertia and Derivative Droop on the DC-link}

To clarify the impact of inertia emulation and damper winding emulation on the VSC DC-link capacitor (see Fig.~\ref{fig:conv_block_diag}) consider the DC-link charge dynamics~\cite{subotic_universal_2024}
    \begin{align*}
    C_{\text{DC}} v^\star_\text{DC} s v_{\text{DC}}(s) = P_{\text{DC}}(s) - P(s), 
    \end{align*}
    where $C_{\text{DC}}$ is the DC-link capacitance, $v_{\text{DC}}$ is the DC-link voltage, and $v^\star_\text{DC}$ is the nominal DC-link voltage. For brevity of the presentation, we assume the DC source implements proportional DC voltage control, i.e., $p_{\text{DC}}(s) = -k_{\text{DC}} v_{\text{DC}}(s)$. This results in
        \begin{align*}
    v_{\text{DC}}(s) = - \frac{1}{k_{\text{DC}}}\frac{1}{\frac{C_{\text{DC}} v^\star_\text{DC}}{k_{\text{DC}}} s + 1}P(s).
    \end{align*}
    In other words, larger DC-link energy storage (i.e., larger $C_{\text{DC}}$) and stiffer DC-link voltage control (i.e., larger $k_{\text{DC}}$) result in less DC voltage fluctuations for a given active power disturbance.
    The relationship between frequency and power for PD droop control is given by
            \begin{align*}
 P(s) = -\frac{T_p s + 1}{m_p\omega_0(1 + \xi_{\text{C}} s)} \omega(s)
    \end{align*}
    and we obtain
    \begin{align*}
    v_{\text{DC}}(s) = \frac{1}{k_{\text{DC}}}\frac{1}{\frac{C_{\text{DC}} v^\star_\text{DC}}{k_{\text{DC}}} s + 1} \frac{T_p s + 1}{m_p\omega_0(1 + \xi_{\text{C}} s)} \omega(s).
    \end{align*}
    Broadly speaking, for a converter that is approximately synchronized to the grid, inertia emulation (i.e., $T_p >0$) acts as high-pass filter on  frequency disturbances as seen from the DC-link voltage and increased inertia emulation (i.e., increasing $T_p$) increases stress on the DC-link capacitor, which corresponds to increased storage requirements. In contrast, damper winding emulation (i.e., $\xi_\text{C} >0$) acts as a lowpass filter on frequency disturbances as seen from the DC-link voltage. Thus, loosely speaking, increased damper winding emulation (i.e., increasing $\xi_\text{C}$) decreases  stress on the DC-link capacitor, decreasing storage requirements.

\subsection{Tuning Derivative Droop}

In practice, $\xi_{\text{C}}$ can be tuned with the Bode plot. For a given bus transfer function $g_{\text{pd}}(s)$ and $R/X$ ratio $\rho$, $M(g_{\text{pd}}, \mu)$ can be determined from the Bode plot of $\frac{1}{\omega}\mu(j\omega)g_{\text{pd}}(j\omega)$ by first identifying $\omega_{\text{c}}$ as the frequency where the phase plot crosses $-90^{\circ}$, and then evaluating the magnitude plot at $\omega_{\text{c}}$ to obtain $M(g_{\text{pd}}, \mu) = \left|\frac{1}{\omega_{\text{c}}}\mu(j\omega_{\text{c}})g_{\text{pd}}(j\omega_{\text{c}})\right|^{-1}$. In this way, $M(g_{\text{pd}}, \mu)$ can be evaluated for a range of $\xi_{\text{C}}$, starting from $\xi_{\text{C}} = (\omega_0\sqrt{1 + \rho^2})^{-1}$. Then, the optimal $\xi_{\text{C}}$ can be selected as the value that generates the largest relative stability margin $M(g_{\text{pd}}, \mu)$. In contrast to other tuning methods, which often require simulating the converter under different operating conditions and subjectively evaluating whether the response seems well-tuned, this method only requires optimizing over a single deterministic metric. Moreover, based on the prior analysis, optimizing over the relative stability margin maximizes the robustness of the converter dynamics, which improves the stability of the system-wide dynamics.

\section{Case Study: VSC Plant}\label{sec:simulation}
We validate the analytical stability conditions by performing an EMT simulation of a modified IEEE 9-bus system where one of the generators is replaced by a VSC plant. We also demonstrate the stability benefits of the reduced-order damper winding emulation control presented in Sec.~\ref{sec:dw_emulation}.

\begin{figure}
    \centering
    
    \begin{subfigure}{\linewidth}
        \centering
        \includegraphics[scale=0.87]{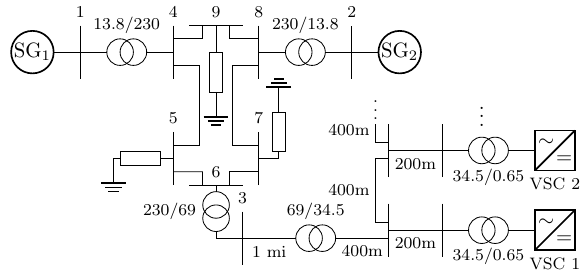}
    \end{subfigure}
    
    \caption{The modified IEEE 9-bus system with a VSC plant added at Bus 3. 
    }\label{fig:sim_setup}
\end{figure}

{\fontfamily{ptm}\selectfont
    \begin{table}
        \caption{Parameters of the simulated system.\label{Table}}
        \begin{minipage}[c]{\linewidth}
        \centering
        \hspace{-5mm}
        \scalebox{0.74}{
            {\centering \renewcommand{\arraystretch}{1.3}
                \begin{tabular}[]{|c|c||c|c||c|c|}
                    \hline
                    \rowcolor{light-gray}
                    \multicolumn{6}{|c|}{System base values}\\
                    \hline
                    $S_\text{b}$ & $100$ MVA & $v_\text{b}$ & $230$ kV& $\omega_\text{b}$ & $2\pi60$ rad/s\\
                    \hline
                    \rowcolor{light-gray}
                    \multicolumn{6}{|c|}{{Synchronous generator (SG)}}\\ \hline
                    $S_r$ & $100$ MVA & $v_r$ & $13.8$ kV& $P^{\star}$ & $75$ W  \\\hline $H$ & $3.7$ s & 
                    $k_g$ & $20$ p.u. & $T_{\text{G}}$ & $3$ s \\\hline $L_{Dd}$ & $0.182$ p.u. &
                    $R_{Dd}$ & $0.0117$ p.u. & $L_{ad}^{\prime\prime}$ & $0.0662$ p.u. \\\hline $L_{aq}^{\prime\prime}$ & $0.1858$ p.u. & & & &  \\ \hline
                    \rowcolor{light-gray}
                    \multicolumn{6}{|c|}{{SG MV/HV transformer}}
                    \\ \hline
                    $S_r$ & $210$ MVA & $v_1$ & $13.8$ kV& $v_2$ & $230$ kV\\\hline
                    $R_1=R_2$ & $0.00338$ p.u. & $L_1=L_2$ & $0.1$ p.u.& $R_\text{m}=L_\text{m}$ & $500$ p.u.\\\hline
                    \rowcolor{light-gray}
                    \multicolumn{6}{|c|}{{VSC Plant MV/HV transformer}}
                    \\ \hline
                    $S_r$ & $210$ MVA & $v_1$ & $69$ kV& $v_2$ & $230$ kV\\\hline
                    $R_1=R_2$ & $0.00338$ p.u. & $L_1=L_2$ & $0.1$ p.u.& $R_\text{m}=L_\text{m}$ & $500$ p.u.\\\hline
                    \rowcolor{light-gray}
                    \multicolumn{6}{|c|}{{VSC Plant MV/MV transformer}}
                    \\ \hline
                    $S_r$ & $120$ MVA & $v_1$ & $34.5$ kV& $v_2$ & $69$ kV\\\hline
                    $R_1=R_2$ & $0.00125$ p.u. & $L_1=L_2$ & $0.05$ p.u.& $R_\text{m}=L_\text{m}$ & $500$ p.u.\\\hline
                    \rowcolor{light-gray}
                    \multicolumn{6}{|c|}{{VSC Plant LV/MV transformer}}
                    \\ \hline
                    $S_r$ & $13.32$ MVA & $v_1$ & $34.5$ kV& $v_2$ & $0.65$ kV\\\hline
                    $R_1=R_2$ & $0.003$ p.u. & $L_1=L_2$ & $0.025$ p.u.& $R_\text{m}=L_\text{m}$ & $500$ p.u.\\\hline
                    \rowcolor{light-gray}
                    \multicolumn{6}{|c|}{{VSC Parameters}}\\\hline
                    $S_r$ & $13.32$ MVA & $v_r$ & $0.65$ kV & $P^{\star}$ & $9.375$ W \\\hline $v_{\text{DC}}^{\star}$ & $2.76$ kV &
                    $k_{\text{DC}}$ & 35.03 & $C_{\text{DC}}$ & $0.008$ F \\\hline 
                    \rowcolor{light-gray}
                    \multicolumn{6}{|c|}{VSC Plant Line Parameters (Pi Line Model)}\\\hline
                    200m $R$ & $66$ $\mu\Omega$/m & 200m $L$ & $0.43$ $\mu$H/m
                    & 200m $C$ & $2.4 \times 10^{-5}$ $\mu$F/m \\\hline
                    400m $R$ & $20$ $\mu\Omega$/m & 400m $L$ & $0.37$ $\mu$H/m
                    & 400m $C$ & $2.8 \times 10^{-5}$ $\mu$F/m \\\hline
                    1 mi $R$ & $0.3076$ $\Omega$ & 1 mi $L$ & $1.8$ mH
                    & 1 mi $C$ & $1.593 \times 10^{-8}$ $\mu$F \\\hline
        \end{tabular}}}
        \end{minipage}
\end{table}}

\subsection{Simulation Setup}
The system (Fig.~\ref{fig:sim_setup}) is simulated in MATLAB Simulink using Simscape Electrical. The simulation lasts for two minutes, with an 0.5 p.u.\ load step at Bus 7 after 30 seconds. The line parameters have varying $R/X$ ratios that, treating stator and transformer impedances as lines, range from $\rho_{\text{min}} = 0.0304$ to $\rho_{\text{max}} = 0.2294$. The synchronous generators are salient-pole machines with droop turbine-governor systems. Each machine also uses an IEEE type ST1A voltage regulator and exciter and a multiband power system stabilizer.

The VSC plant is composed of eight two-level converters, each represented by an averaged VSC model. Each converter is attached to its own DC generation, which is represented as a controlled current source with a current $i_{\text{DC}}$ depends on its voltage $v_{\text{DC}}$ such that $i_{\text{DC}} = P^{\star}/v_{\text{DC}}^{\star} + k_{\text{DC}}(v_{\text{DC}}^{\star} - v_{\text{DC}})$, where $P_{\text{DC}}^{\star}$ and $v_{\text{DC}}^{\star}$ are the active power and DC voltage setpoints and $k_{\text{DC}} \in \R_{\geq 0}$ represents the  generation response to DC voltage deviations. This model mimics the small-signal response of curtailed solar PV and battery energy storage systems. The outer converter controls are implemented using nested proportional-integral current and voltage loops. The parameters for the VSC plant lines and transformers are adapted from~\cite{ramasubramanian_gfm_2025}. The complete system paramters are listed in Table~\ref{Table}. 

\subsection{Converter Implementation}
\begin{figure}
    \centering
    \begin{subfigure}{\linewidth}
        \centering
        \includegraphics{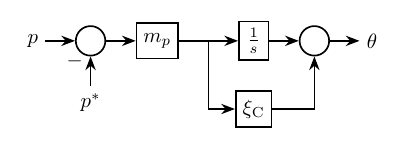}
        \subcaption{Implementation when $T_p = 0$.}\label{fig:impl_tp_0}
    \end{subfigure}
    \begin{subfigure}{\linewidth}
        \includegraphics[width=\linewidth]{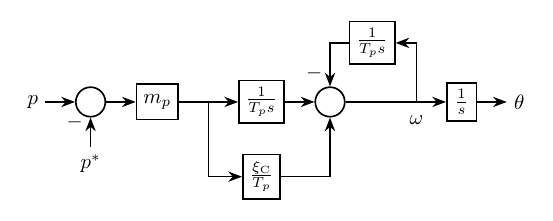}
        \subcaption{Implementation when $T_p > 0$.}\label{fig:control_implementation_tp}
    \end{subfigure}
    \caption{Realizable implementations of PD droop control.}\label{fig:conv_outer_control}
\end{figure}
The converters are modeled using grid-forming control with inner loops for both the voltage and current. The inner control loops are model using standard decoupled proportional-integral control in $dq$ frame. The converters are represented by an averaged model, which means the simulation does not include the effects of switching ripple. In practice, modern power electronic converters achieve switching frequencies well above the frequency ranges of interest in the outer loop controls, and by timescale separation we can thus neglect any interactions between the switching frequencies and the slower timescales discussed here.

We compare two different outer loop controls: conventional and PD droop. For both controls, we use a droop coefficient of $m_p = 0.05$ and an active power filter time constant of $T_p = 3$s. For the PD droop control, we use $\xi_{\text{C}} = 0.005$. Note that PD droop control does not require a derivative term in its implementation (Fig.~\ref{fig:control_implementation_tp}), avoiding the increased noise amplification traditionally associated with realizable derivatives in controls. Even if $T_p = 0$, the control is still realizable since the VSC only requires an angle reference, not a frequency reference (see Fig.~\ref{fig:impl_tp_0}). For the voltage magnitude reference, we use standard $Q-V$ droop control for both controls.

\subsection{Results with Conventional Droop Control}
First, we analyze the system if all converters implement conventional droop control. The outer controls are modeled by $g_{\text{dr}}(s)$ with $m_p = 0.05$ and $T_p = 3$s, which satisfy the lower bound given by~\eqref{eq:T_lower_bound} and fall within the range of commonly suggested parameters for battery energy storage systems. Note that Assumption~\ref{assumption:r/x} is not satisfied for this system, and the line $R/X$ ratios range from $\rho_{\text{min}}$ to $\rho_{\text{max}}$. However, we can still apply the analytical stability criteria by analyzing stability for both $\mu(s)$ defined with $\rho_{\text{min}}$ and $\mu(s)$ defined with $\rho_{\text{max}}$. 

\begin{figure}
    \centering
    \includegraphics{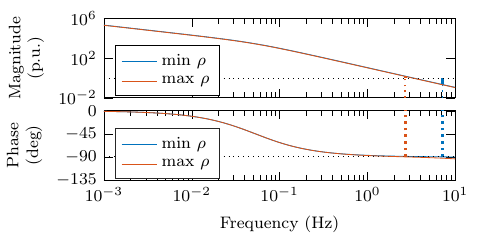} 
    \caption{Bode plot of $\frac{1}{\omega}\mu(j\omega)g'(j\omega)$ for VSC 4 with bus dynamics $g_{\text{dr}}(s)$ where $m_p = 5\%$ and $T = 3$. The plot is shown for $\mu(s)$ with $R/X$ ratios $0.0304$ and $0.2294$, the minimum and maximum system values. The vertical dotted lines correspond to $\omega_\text{c}$ for each transfer function, and the solid vertical lines on the magnitude plot denote the magnitude at $\omega_\text{c}$. For the maximum $\rho$, the gain at $\omega_\text{c}$ is greater than 1, indicating potential instability.
    }\label{fig:analysis_T3kd0_vsc}
\end{figure}

Applying the individual analytical requirements to each converter, we see that this control results in potentially unstable behavior (Fig.~\ref{fig:analysis_T3kd0_vsc}). In particular, for higher $R/X$ ratios, $M(g_{\text{dr}}, \mu) < \gamma_n$ is not satisfied for nodes $n$ within the VSC plant. Moreover, while the condition appears satisfied for lower $R/X$ ratios (i.e., $\frac{1}{\omega}|\mu(j\omega)g'(j\omega)| < 1$ when $\angle \mu(j\omega)g'(j\omega) \notin (-90^{\circ}, 90^{\circ})$), the phase plot for $\frac{1}{\omega}\mu(j\omega)g'_{\text{dr}}(j\omega)$ is very flat around $-90^{\circ}$. This means that slight changes to the phase response could reduce $\omega_\text{c}^{\text{dr}}$ to an unstable value, indicating a lack of robustness to unmodeled dynamics. 

In EMT simulation, these control values lead to an unstable system response (Fig.~\ref{fig:EMT_sys}) and a growing oscillation between the frequency and power responses of VSCs within the plant (Fig.~\ref{fig:EMT_plant_osc} and \ref{fig:EMT_plant_pow}). The frequency oscillates at approximately 1.7 Hz, which is close to the instability around 2.7 Hz suggested by the Bode plot in Fig.~\ref{fig:analysis_T3kd0_vsc}.
We speculate that the frequency discrepancy is due to the sensitivity of the crossover frequency of VSC droop control to dynamics in the simulated system that are not captured in the small-signal model. 

\subsection{Results with PD Damper Winding Emulation}
\begin{figure}
    \centering
    \includegraphics{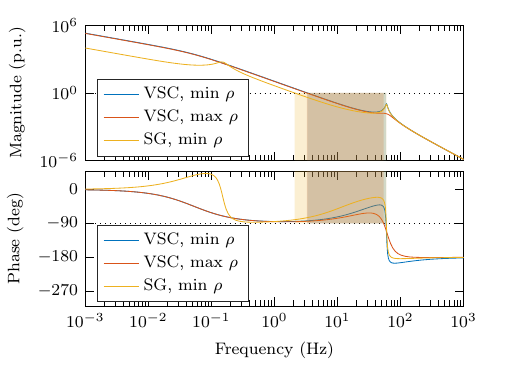} 
    \caption{Bode plot of $\frac{1}{\omega}\mu(j\omega)g'(j\omega)$ is shown for a synchronous machine and a converter in the modified 9-bus system when the converter dynamics are given by $g_{\text{pd}}(s)$ with $m_p = 5\%$, $\xi_{\text{C}} = 0.005$ and $T = 3$. The plot for the inverter is shown for the minimum and maximum system $R/X$ ratios. Each Bode plot is shaded with its corresponding color over the frequency range where $\Re(\frac{1}{\omega}\mu(j\omega)g'(j\omega)) > 0$ and $|\frac{1}{\omega}\mu(j\omega)g'(j\omega)| < 1$. The overlap of these shaded regions, shown in brown, corresponds to Region 2 in Condition~\ref{cond:interpretable}.}\label{fig:analysis_T3kd0.005}
\end{figure}

Next, we control each converter with PD damper winding emulation $g_{\text{pd}}(s)$ with $\xi_{\text{C}} = 0.005$, where again $m_p = 0.05$ and $T_p = 3$s. This increases $\omega_{\text{c}}$ for the converters to near line frequency, where the gain of the frequency response is much smaller. As a result, the converters in the VSC plant have a much larger stability margin across the range of $R/X$ values. Plotting the frequency response of the converters and the machines on the same axes (Fig.~\ref{fig:analysis_T3kd0.005}) reveals there is now a significant frequency range (3.3 Hz -- 54.0 Hz) for which the converters and machines all have both positive real part and gain less than one. Therefore, we can certify interoperability via Condition~\ref{cond:interpretable} with $\omega_1 = 3.3$ Hz and $\omega_2 = 54$ Hz.

\begin{figure*}
    \centering

    \begin{subfigure}{\textwidth}
        \centering
        \includegraphics[width=\textwidth]{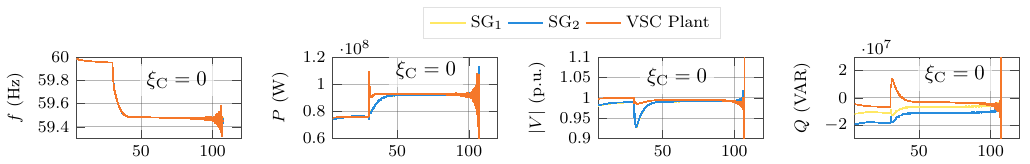}

        \includegraphics[width=\textwidth]{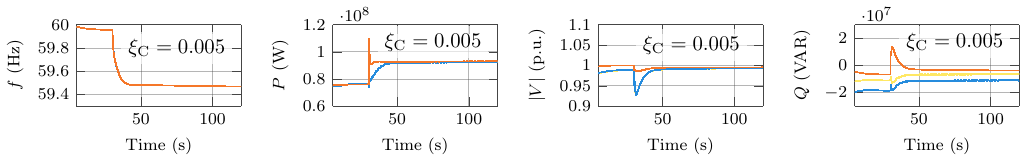}
        \caption{Simulation results for the synchronous generators and the VSC plant for the case where $\xi_{\text{C}} = 0$ (above) and $\xi_{\text{C}} = 0.005$ (below).}\label{fig:EMT_sys}
    \end{subfigure}
    
    \vspace{1em}\begin{subfigure}{\textwidth}
        \centering
        \includegraphics[width=\textwidth]{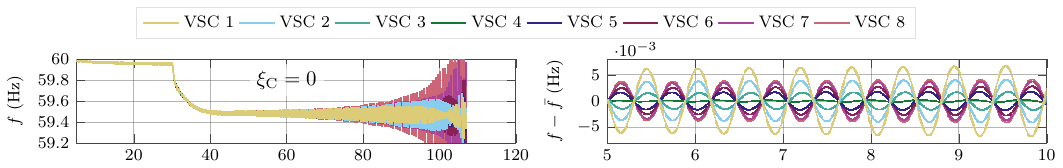}

        \includegraphics[width=\textwidth]{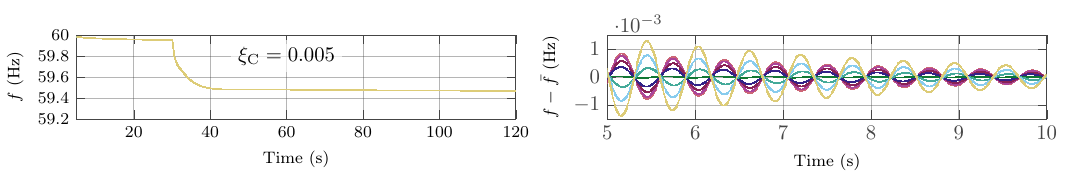}
        \caption{Converter-level frequency responses for each VSC in the plant for $\xi_{\text{C}} = 0$ (above) and $\xi_{\text{C}} = 0.005$ (below). On the left is the full frequency response, and on the right is a zoomed-in plot of the difference between each frequency and the average VSC frequency $\bar{f}$.}\label{fig:EMT_plant_osc}
    \end{subfigure}

    \vspace{1em}\begin{subfigure}{\textwidth}
        \centering
        \includegraphics[width=\textwidth]{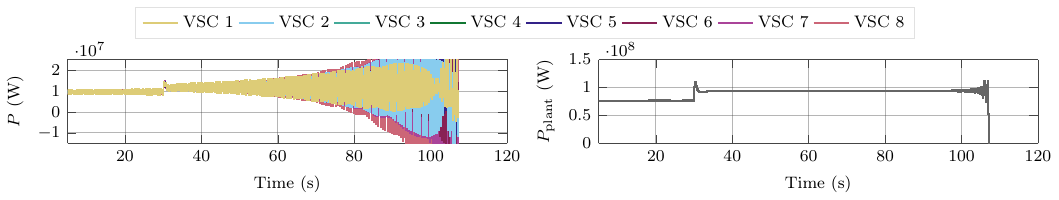}
        \caption{The power response for the individual VSCs (left) and the total VSC plant (right) for the simulation with $\xi_{\text{C}} = 0$. The power oscillation within the plant starts early and grows large quickly, but because the VSCs are oscillating against each other this oscillation is not visible in the total plant power output $P_{\text{plant}}$ until later in the simulation.}\label{fig:EMT_plant_pow}
    \end{subfigure}

    \caption{EMT simulation results for the modified 9-bus system with a VSC plant using conventional droop control (i.e., $\xi_{\text{C}} = 0$) and PD droop control (i.e., $\xi_{\text{C}} = 0.005$). Using conventional droop control, an unstable frequency oscillation occurs in the VSC plant. In contrast, the PD droop control is able to damp that oscillation and maintain stability of the system.}\label{fig:sim_results}
\end{figure*}

The improved analytical stability margin translates to improved performance in simulation. The system is able to maintain stability throughout the full 2 minute simulation (Fig.~\ref{fig:EMT_sys}). Notably, the frequency oscillations observed within the VSC plant with conventional droop are well-damped with the PD droop control (Fig.~\ref{fig:EMT_plant_osc}). This enhanced frequency convergence matches the effect of incorporating the reduced-order damper winding model into the synchronous machine model for multi-machine systems demonstrated in Fig.~\ref{fig:freq_model_dw}.

\section{Conclusion}
This paper presented a reduced-order model of damper windings that improved the small-signal model of synchronous machines. The paper also introduced a set of decentralized conditions for ensuring the small-signal frequency stability of a system with heterogeneous bus dynamics and homogeneous line dynamics. With these stability conditions, it was shown that incorporating the reduced-order damper winding model improves the stability margin of synchronous machines under common operating conditions. Further, this paper showed that PD droop control for converters can be interpreted as emulating the synchronizing effect of damper windings, and the presented stability conditions were used to demonstrate the stability benefits of PD droop control over conventional droop control. Finally, the advantages of PD droop control were demonstrated in an EMT simulation.

The analysis presented here only demonstrated small-signal stability, so while it is a good approximation of transmission line dynamics, it cannot capture the full nonlinear dynamics of the power system. Additionally, this model neglects the AC voltage magnitude and reactive power dynamics. This paper also did not explicitly model the DC voltage dynamics and excluded high voltage DC lines. Developing stability criteria for the dynamics of AC voltage magnitude and DC voltage is seen as an interesting direction of future work. Other areas of future work include incorporating robustness margins into the stability criteria to account for model inaccuracies (e.g., nonlinearity and heterogeneous $R/X$ ratios). 


{\appendix[Proof of Lemmas]}
\begin{proof}[Proof of Lemma~\ref{lemma:coherent}]
    To show that the closed-loop system will have a bounded response to disturbances $s \in S_{\delta}$, we show that the closed-loop transfer function $H(s) \coloneqq {\left(I + G'(s)\frac{\mu(s)}{s}L'\right)}^{-1}G'(s)$ will have bounded magnitude for $s \in S_{\delta}$. By~\cite[Thm 3]{min_frequency_2025}, Condition~\ref{cond:stable_steady_state} guarantees that the system will converge to $\bar{g}(s)$ in steady state. Moreover, Condition~\ref{cond:stable_steady_state}.\ref{cond:bounded_coherent_dyn} ensures that the synchronous dynamics will have bounded magnitude over $s \in S_{\delta}$. We can further assert that for $s \in S_{\delta}$, $|H(s)|$ will be a finite distance from $|\bar{g}(s)|$~\cite[Lemma 1]{min_frequency_2025}, ensuring $|H(s)|$ is bounded over $s \in S_{\delta}$.
\end{proof}

\begin{proof}[Proof of Lemma~\ref{lemma:interoperable}]
    To prove stability over $s \in \overline{\C}_+ \setminus S_{\delta}$, we adapt the arguments used in~\cite{pates_robust_2019} and~\cite[Thm 1]{bui_input-output_2024}. Condition~\ref{cond:formal_transient_stability}.\ref{cond:no_transient_poles} implies that each $g_n'(s)$ will be analytic in the interior of $\overline{\C}_+ \setminus S_{\delta}$, so by the maximum modulus principle, we can show $|H(s)|$ is bounded over $\overline{\C}_+ \setminus S_{\delta}$ by showing that $|H(s)|$ is bounded over the boundary of the set. 
Because by Condition~\ref{cond:formal_transient_stability}.\ref{cond:no_transient_poles} $|G'(s)|$ is bounded in $\overline{\C}_+ \setminus S_{\delta}$, $|H(s)|$ is bounded on $\overline{\C}_+ \setminus S_{\delta}$ if and only if zero is not an eigenvalue of $I + G'(s)\frac{\mu(s)}{s}L'$ for all $s$ on the boundary of $\overline{\C}_+ \setminus S_{\delta}$. 
Applying the logic of ~\cite[Thm. 1]{pates_robust_2019}, it is sufficient to show that $0 \notin \text{Co}\left(\left\{1 + k\frac{\mu(s)}{s}g_n'(s)\: : \: n \in \mathcal{N},\, 0 \leq k \leq 1\right\}\right)$ for all $s$ on the boundary of $\overline{\C}_+ \setminus S_{\delta}$, where $\text{Co}$ denotes the convex hull of a set. Applying the separating hyperplane principle, this is satisfied for a given $s$ if and only if there exists some angle $\phi \in [0, \pi/2)$ such that 
$\Re\left(e^{j\phi}\left(1 + \frac{\mu(s)}{s}g_n'(s)\right)\right) > 0$
for all $n \in \mathcal{N}$. Condition~\ref{cond:formal_transient_stability}.\ref{cond:interoperability} guarantees that this holds for all $s$ on the boundary of $\overline{\C}_+ \setminus S_{\delta}$.
\end{proof}

\bibliographystyle{IEEEtran}
\bibliography{IEEEabrv,ref}


 





\end{document}